\documentclass[lettersize,journal]{IEEEtran}
\usepackage{amsmath,amsfonts,amsthm}
\usepackage{algorithmic}
\usepackage{array}
\usepackage[caption=true,font=small]{subfig}
\usepackage{textcomp}
\usepackage{stfloats}
\usepackage{url}
\usepackage{verbatim}
\usepackage{graphicx}
\usepackage{balance}
\usepackage{xcolor}

\ifodd 0

\newcommand{\rev}[1]{{\color{blue}#1}}

\newcommand{\com}[1]{{\color{red}\textbf{(#1)}}}%comment of the text
\else

\newcommand{\rev}[1]{#1}

\newcommand{\com}[1]{}
\fi

\newtheorem{example}{Example}
\newtheorem{theorem}{Theorem}
\newtheorem{proposition}[theorem]{Proposition}%[section]
\newtheorem{assumption}{Assumption}
\newtheorem{lemma}{Lemma}%[section]

\DeclareMathOperator*{\argsup}{arg\,sup}

\begin{document}
\title{Incentivizing Secure Software Development: the Role of Voluntary Audit and Liability Waiver\thanks{A preliminary version of this paper is appearing in the Annual IEEE Control \& Decision Conference 2024.}}
\author{Ziyuan Huang, Gergely Bicz{\'o}k, Mingyan Liu 
\thanks{Ziyuan Huang and Mingyan Liu are with the Electrical and Computer Engineering Department, University of Michigan, 1301 Beal Avenue, Ann Arbor, MI 48109, USA; e-mail: \texttt{\{ziyuanh, mingyan\} @umich.edu}; Gergely Bicz{\'o}k is with the CrySyS Lab, Department of Networked Systems and Services, Budapest University of Technology and Economics, 1111 Budapest, Műegyetem rkp. 3, Hungary; e-mail: \texttt{biczok@crysys.hu}. Gergely Bicz{\'o}k was also with the University of Michigan when this research was done.}}

% \markboth{Journal of \LaTeX\ Class Files,~Vol.~18, No.~9, September~2020}%
% {How to Use the IEEEtran \LaTeX \ Templates}

\maketitle

\begin{abstract}
Misaligned incentives in secure software development have long been the focus of research in the economics of security. Product liability, a powerful legal framework in other industries, has been largely ineffective for software products until recent times. However, the rapid regulatory responses to recent global cyber attacks by both the United States and the European Union, together with the (relative) success of the General Data Protection Regulation in defining both duty and standard of care for software vendors, may just enable regulators to use liability to re-align incentives for the benefit of the digital society. Specifically, the recently proposed United States National Cybersecurity Strategy suggests shifting responsibility for cyber incidents back to software vendors. In doing so, the strategy also puts forward the concept of the liability waiver: if a software company voluntarily undergoes and passes an IT security audit, its future product liability is (fully or partially) waived.

In this paper, we analyze this audit scenario from the perspective of the software vendor and the auditor, respectively. From the vendor's point of view, this is formulated as a sequential decision problem: a vendor with a product or process needs to pass a mandatory audit in order to be able to release the product onto the market; it is allowed to go through the audit repeatedly, and thus the vendor needs to determine what level of effort to put into the product (e.g., to enhance its security) following each failed audit. We examine the vendor's optimal decision process and fully characterize its properties under mild technical assumptions.  From the auditor's point of view, we examine what happens if the audit is optional; in particular, we seek to answer the question of what type of audit might be the most effective in incentivizing voluntary participation and, at the same time, more desirable effort from the vendor.
%We propose a mechanism where a software vendor should first undergo a repeated auditing process in each stage of which the vendor decides whether to quit early or stay with additional security investment. 
We show that the optimal strategy for an opt-in vendor is to never quit and to exert cumulative investments in either a ``one-and-done'' or ``incremental'' manner. We also show that to incentivize participation and high effort from a vendor, a desirable audit rule should be highly accurate but not very strict. 
We also showed how a dynamic audit setting can be exploited to increase the vendor's incentivizable investment under linear audit rules.

\end{abstract}


\section{Introduction}
\label{sec:intro}

Making software products more secure is arguably one of the most important elements in securing our overall computer and information network ecosystem. It has also been one of the most challenging due in no small part to a sequence of misaligned incentives. For one, security features in a software product can be hard to monetize (even when they are noticeable), and thus, spending resources to improve the security in software production may not lead to immediate or substantial returns on investment. This is exacerbated by the fact that software markets usually reward first-movers that release new functional features as quickly as possible, resulting in little to no security testing; hence the mantra: ``we'll ship it on Tuesday and get it right by version 3''~\cite{DBLP:conf/acsac/Anderson01}.  Secondly, while vulnerabilities in software products can lead to substantial costs to a vendor (incurred in developing patches, for instance), there is significant uncertainty on whether certain costs will materialize; a large number of vulnerabilities are never discovered or publicly disclosed, and even among those discovered, a majority of them are never exploited. Therefore, rather than fixing the vulnerabilities, they are often rolled up in the development of newer versions of the software. This means there is less motivation on the vendor's part to try to minimize vulnerabilities a priori.  Perhaps most importantly, a software vendor's exposure to potential security risk is limited by the fact that the vast majority of the cost incurred in a security incident is borne by the consumer/user/buyer of the software, not the producer, instantiating a lax approach to risk mitigation owing to moral hazard~\cite{DBLP:conf/acsac/Anderson01}.

Modern liability frameworks, emerging from early 20th-century case law~\cite{macpherson_1916}, aim at establishing legal obligations for individuals and organizations to assume responsibility for their actions, particularly when such actions result in harm or damage to others. Liability encompasses various legal principles and frameworks that determine when (duty of care~\cite{donoghue_1932}) and to what extent (standard of care~\cite{us_1947}) one party may be held accountable for the consequences of their behavior. Historically, software companies frequently avoided product liability using a combination of legal gray zones and disclaimers in end-user license agreements (EULA), capitalizing on the broad interpretation of acceptable user risk. However, the General European Data Protection Regulation (GDPR~\cite{GDPR2016a}) leveled the playing field by defining both duty and standard of care, leading to substantial fines for mishandling personal data and prompting a reassessment of cybersecurity investments. In addition, recent global cyberattacks such as the 2017 NotPetya and the 2021 SolarWinds incidents triggered rapid regulatory responses in the US and the EU, mandating secure software development, compliance, and supply chain security to overcome information asymmetry and lack of expertise for end-users, re-assigning liability back to software vendors. At the same time, the insurance sector has been grappling with the insurability of cyber risks, particularly in critical infrastructure, following severe cybersecurity incidents and ongoing armed conflicts~\cite{wef_2023}. The insurance industry is scrambling to establish baseline scenarios for industrial control systems~\cite{dagstuhl_2022}, as the potential for systemic cyber risks and catastrophic losses may shift responsibility to governments as insurers of last resort~\cite{cummins2006should}. This motivated a series of national security policy directives that aim to allow liability claims against insecure software products produced by software vendors. In the United States, the US National Cybersecurity Strategy~\cite{national_cybersecurity_strategy}, released in April 2023, has put forward a liability waiver mechanism tied to government-mandated security audits, serving as a financial incentive for software companies to improve their product security practices. While the expected impact and an efficient manner of implementation are still heavily debated in the US~\cite{lawfare_Dempsy24}, the EU has already passed an elaborate web of cybersecurity and liability regulations (a combination of the Cybersecurity Act~\cite{CSA2019}, the Cyber Resilience Act~\cite{CRA2022proposal}, and the new Product Liability Directive~\cite{PLD2022proposal}, among others) effectuating mandatory certification (by way of third-party audits for higher levels of assurance) for software products.
%Gergely: add EU - DONE
%Gergely: add some references for support claims - DONE

%%%%%%%%%%%%%%%%%%%
\rev{In this work, we examine what happens if a (government) agency offers optional (and free) product security audits and, for those who pass the audit, fully or partially waives their liability associated with software security.  This is studied from both the vendor's and the auditor's perspectives, respectively. From the vendor's perspective, we examine its best strategy in terms of security investment in its product development if it decides to go through the audit. This is formulated as a sequential decision problem faced by the vendor: it needs a successful audit in order to bring the product to market and can go through the audit repeatedly following each failure. The audit mechanism is known to the vendor, and it thus needs to determine what level of effort to put into the product (e.g., to enhance its security/quality/performance) following each failure. The audit is assumed to be informative but imperfect, with some built-in randomness, i.e., the outcome contains both false positive and false negative results. We fully characterize the properties of the vendor's optimal decisions under mild technical assumptions.  
We then study the problem from the auditor's perspective, i.e., 
%We examine this problem in two settings: when the audit is mandatory and when the audit is optional; in the latter case, 
how to incentivize the vendor to voluntarily subject itself to the audit. In particular, we examine what type of audit might be the most effective in
incentivizing voluntary participation and, at the same time, more desirable effort from the vendor. 
}

%This study is primarily motivated by the pressing issue of software security facing the software industry. Making software products more secure is arguably one of the most important and challenging elements in securing our overall computer and information network ecosystem. For one, security features in a software product can be hard to monetize; for another, a software vendor's exposure to potential security risk is limited by the fact that the vast majority of the cost incurred in a security incident is borne by the consumer of the software. All these facts diminish the vendor's motivation to minimize software vulnerabilities a priori, leading to the problem of moral hazard \cite{DBLP:conf/acsac/Anderson01}. This also motivates a series of national security policy directives that aim to allow liability claims against insecure software products. Specifically, the United States National Cybersecurity Strategy (USNCS), released in April 2023\cite{national_cybersecurity_strategy}, has proposed the possibility of a liability waiver mechanism tied to government-mandated security audits, serving as a financial incentive for software companies to improve their product security practices.

Existing literature on audit incentives assumes strategic auditors. 
\cite{john1985,Ella1992,finley1994,ben2015auditing} model the auditing process with simultaneous-move games between an auditor and an auditee. The auditor inputs costly auditing design efforts aiming to achieve the highest accuracy net design costs, while the auditee tries to pass the audit with minimum investment costs. A similar approach was adopted by \cite{patterson_audit_2003,brown_incentives_2007} on carefully designed finite sequential audit games. The auditing problem has also been studied in the insurance literature \cite{khalili_designing_2018,khalili_embracing_2019,Nugrahanti_2023} where the insurer acts as the auditor trying to maximize its profit less the audit cost.
Our model relaxes the budget requirement on auditing efforts, enabling indefinitely repeated interactions between the auditor and the auditee (vendor). We also do not assume strategic auditors in this work so as to focus exclusively on understanding the impact of the audit structure. \rev{To the best of our knowledge, this paper is the first attempt at modeling enforced repeated audits in the software industry, which are}  more suitably implemented by regulatory authorities such as governments 
%to ensure pure benignity of vendors, 
as suggested by the US National Cybersecurity Strategy~\cite{national_cybersecurity_strategy}, in contrast to profit-maximizing audit models adopted by the vast insurance and accounting literature.

Our goal in this paper is to understand: (1) from the vendor's perspective, what is the optimal sequence of investment under audit, (2) from the auditor's perspective, how to maximally incentivize vendors to opt into such an audit. Our main findings are the following: 
\begin{itemize}
\item \rev{In the static audit case, where the same test is administered independently after each failure,} the optimal policy for the vendor under audit is, in general, non-unique but enjoys some very interesting properties.  An optimal policy falls into two broad categories: the ``one-and-done'' type and the ``incremental'' type (Section \ref{sec:opt-in}). Under the first type, the vendor invests in one installment at the beginning of the process, an amount from a well-defined optimal set, prior to the initial audit, and then waits to pass the audit, even if it takes multiple rounds.  %For some problem instances, this initial investment could be quite low (in which case the vendor simply waits for luck to carry it through the audit). 
Under the second type, the vendor invests multiple times with each amount from the same optimal set, but the timing of these investments can be arbitrary. 
%\item Under either type of strategy, given the probabilistic nature of the audit outcome, an audit could clear a product with very low security investment for market entry, which is undesirable from a social welfare standpoint.  The auditor can mitigate this by making the audit more accurate (lower noise) but easier (lower threshold for passing). 

%\item We show how this audit mechanism can be related to a liability waiver insurance policy, and how it reshapes the vendor's risk perception. 

\item We show how the audit quality (its accuracy and hardness) influences the vendor's participation incentive and how to adjust these parameters to increase participation (Section \ref{sec:audit-quality}). 

\item \rev{
In the dynamic audit case, where a different test may be used depending on the history of past failures, we show that a finite-step audit (a finite sequence of tests followed by an independent repetition of the last one indefinitely) can reasonably approximate any general dynamic audit. Of particular interest is the special case of a two-step audit, where a second test is administered repeatedly and indefinitely following the failure to pass a first test. We show that for linear-form audits, an easier audit followed by an infinite sequence of harder ones can incentivize higher investments than a comparable static audit, while a hard audit followed by an infinite sequence of easier ones cannot (Section \ref{sec:dynamic}).
 } 
\end{itemize}

The remainder of the paper is structured as follows.  
%In Section \ref{sec:related} we review related literature.  
Section \ref{sec:problem-setup} presents an optimal stopping time model to capture the decision-making process of a vendor and some key properties of this problem.  Section \ref{sec:opt-in} fully characterizes the vendor's optimal strategy if it decides to participate in the audit mechanism.  Section \ref{sec:audit-quality} examines how the audit parameters influence the participation incentive of a vendor. \rev{Section \ref{sec:dynamic} shows how the auditor can further improve the efficacy of the audit by adopting a dynamic auditing method, whereby successive audits may depend on the history of the audit (e.g., how many times the vendor has failed in the past).} 
Section \ref{sec:discuss} concludes the paper and discusses several potential extensions.

\section{Model and Preliminaries}\label{sec:problem-setup}

The basic problem consists of a neutral auditor with a pre-determined, publicly known audit rule and a utility-maximizing (software) vendor (also referred to as the developer or producer) responding optimally to the audit rule.  The process plays out in discrete time and potentially over multiple periods, as the vendor may need to be audited repeatedly \rev{(via potentially different tests)} in order to pass. Below, we describe the sequential decision problem and related assumptions, 
%in Section \ref{sec:problem} and then briefly introduced the 
followed by the solution approach. 
%, i.e., the discounted-reward Markov Decision Process (MDP) formulated as an optimal stopping problem, in Section \ref{sec:method}.

\subsection{Problem Description}

Consider the scenario where a vendor has to satisfy a certain auditing requirement to enter the market. \rev{The audit consists of a sequence of tests, which are executed successively and} stop only when the vendor passes any test or quits the market entirely. The rationale for the latter assumption is that a product with known defects/vulnerabilities should not be allowed to enter the market. \rev{We say the vendor passes the audit when it passes any test in the sequence of tests constituting the audit.}
The vendor is utility maximizing and potentially risk averse, optimizing over its (successive)  investment/effort levels (e.g., to enhance the security of the product), whereas the auditor is not. We shall assume that the audit service is free but not perfect in its accuracy, i.e., the outcome of each test may contain both false positives and false negatives. More importantly, for the most part, we will assume that successive tests in the audit are independent of each other. \rev{This assumption is relaxed in Section \ref{sec:dynamic}, where the auditor adopts a dynamic auditing method that takes into account the history of past test outcomes.}

% The availability of such an audit service may be viewed as a type of mechanism. 
% The central questions we seek to answer include whether there is incentive for a vendor to participate in such a mechanism, and whether it can induce better/higher security effort from the vendor in its software production.  
%%%%%%%%

\subsubsection*{The Vendor}\label{sec:problem}
From the vendor's point of view, the auditing process can be modeled as an infinite-horizon dynamic system in discrete time. %(or a single-player dynamic game).  
Let each time step of this process be indexed by $t\in\{0,1,\dots\}$. The vendor must determine and commit to an investment $x_0\in\mathbb{R}_+$ at time $t=0^{+}$. 
The value $x_0$ is the private information of the vendor; however, since its utility function is assumed public knowledge, the vendor's optimal strategy, including the value $x_0$, is ultimately known to the auditor (i.e., the latter can simply follow the same computation)\footnote{This point is not particularly relevant in the present setting since the auditor is not assumed to be strategic or even profit-maximizing; but it would become relevant in various extensions of the basic model.}. %\com{Perhaps we can add: This point is not particularly relevant in the present setting since the auditor is not assumed to be strategic or even profit-maximizing; but it would become relevant in various extensions of the basic model.}
%as the investment must be fully implemented rather than simply proposed for the auditor's evaluation. 
% Equivalently, it can also be interpreted as the vendor receiving an instantaneous continuation reward of $r_0=-C_X(x_0)$ for continuing the process. 
The effort $x_0$ goes into the software development over the first time step, incurring a cost of $c_0=C_X(x_0)$ by time $t=1^{-}$, when the product is submitted for \rev{the first test in the audit}, with its outcome revealed at time $t=1$. $C_X(\cdot)$ is assumed positive and continuous.  
% \rev{An instantaneous cost of $c_0=C_X(x_0)$ is incurred at the completion of the product at $t=1^-$.}
%The process then transitions to stage $t=1$, where the (random) auditing outcome is revealed (see the audit rule in the next subsection). 
Test outcome at time $t$ is denoted by $s_t\in\{0,1\}$.
If the outcome is positive/successful, i.e., $s_1=1$, the process terminates: the vendor is granted market entry, earning a reward/revenue $r_1$, assumed to be a constant $R$ at time $t=1^{+}$.  
%That is, the vendor only bears the security investment cost and is waived of potential software liability. 
This is also considered the terminal reward for passing the entire audit
% , denoted as $r_P:=R$ when achieved at time $t^{+}$ 
with discounting applied through the utility function given shortly below.

\begin{figure}[!htbp]
    \centering
    \includegraphics[width=\linewidth]{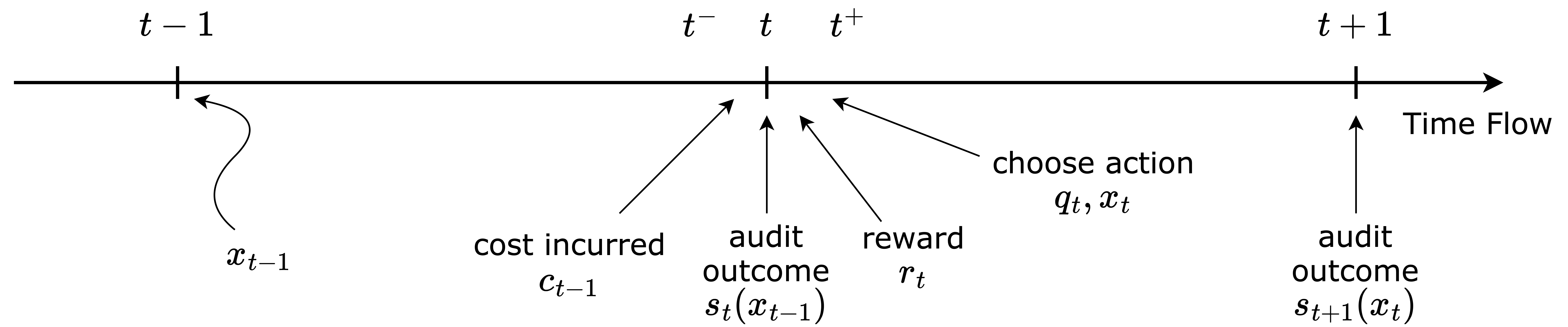}
    \caption{Timeline of the decision process. 
    % Variables arrive successively as time flows from left to right. 
    % Blue blocks denote the binary auditing results $s_t$, green blocks denote rewards $r_t$, red blocks denote instantaneous costs $c_t$, and yellow blocks denote the vendor's decisions. $q_t$ is a binary decision output of quitting at time $t$ with $q_t=1$ indicating a quit and $x_t$ is the new investment level at time $t$. Since we assume the process has not stopped until $t+1$, these variables should satisfy $r_1=\cdots=r_{t+1}=0$ and $q_1=\cdots=q_{t+1}=0$.
    }
    \label{fig:timing}
\end{figure}

If the test outcome is negative/fail, i.e., $s_{1}=0$, then the vendor is temporarily denied market entry. It can either choose to quit the process (exit the market) at time $t=1^{+}$, thereby receiving zero revenue but incurring no further cost, or opt for re-auditing. In the latter case, the vendor must decide a new (cumulative) effort level $x_1\in[x_{0},\infty)$ at time $t=1^{+}$, thereby committing to an {\em additional} investment of $x_1-x_{0}$ over the next time step.  %\rev{The implementation of the additional security investment is time-consuming and persists throughout the time step.} 
This incurs a cost of $c_1=C_X(x_1)-C_X(x_{0})$ at time $t=2^{-}$.
% equivalently, this may be viewed as a continuation reward of $r_1=-C_X(x_1)+C_X(x_0)$. 
The process then proceeds to stage $t=2$ and repeats indefinitely until the vendor either successfully passes the audit or quits. 
% We will use $\rev{r_t=}r_Q:=0$ to denote the terminal reward for quitting when decided at time $t^{+}$.  
For any non-terminal stage, we assume it generates a reward of $r_t=0$.  
Let $q_t$ denote the binary continuation decision with $q_t=1$ indicating a quit.
The sequence of actions and decisions is illustrated in Fig. \ref{fig:timing}, assuming the process has not stopped by $t+1$. 

% \rev{As is shown in Fig. \ref{fig:timing}, there also exists a discount factor $\alpha\in(0,1)$ that discounts future rewards/costs in the vendor's utility. It means that the vendor pays more attention to immediate gains/losses compared to future ones. Starting from $t=1$, either cost or reward incurred at time $t$, i.e., in either $t^{-}$ or $t^{+}$, is discounted in magnitude by $\alpha^{t-1}$. The vendor's utility is the sum over the discounted rewards minus the sum over discounted costs.}

% \rev{We highlight the distinctions between decision periods and discount regions. The former proceeds with the vendor observing the audit outcome $s_t$ and associated reward $r_t$, then choosing the actions $q_t$ and $x_t$, and waiting till the eve of the next time step when the action cost is materialized and the new audit application is submitted. Variables involved in the same decision period have the same time index $t$ as in Fig. \ref{fig:timing}. On the contrary, the latter, as explained above, is defined for each sharp timestamp $t$. All rewards/costs received from near the same timestamp belong to the same discount region and are discounted with the same value. Discount region at each time $t$ usually contains the security investment cost from $t-1$ (actually incurred at $t^{-}$) and the reward from $t$ (actually received at $t^{+}$).}

Formally, define $\underline{x}:=\{x_t\}_{t=0}^\infty$ as the increasing sequence of total (cumulative) investments by the vendor at time $t^{+}$.
Let $\tau_s\in\{1,2,\dots\}$ denote the quitting time and $\tau_a\in\{1,2,\dots\}$ the time at which the vendor first passes the audit; both are in general random (stopping) times of the processes $\{q_t\}_{t\geq0}$ and $\{s_t\}_{t\geq0}$ respectively. 
% Let $\alpha\in(0,1)$ be the discount factor that discounts future rewards/costs. % so that the vendor attaches higher value to immediate rewards. 
%Assume all information, including the history investment decisions and instantaneous or terminal rewards, are available to the vendor. Therefore, 
The vendor's utility can be written as follows, given its decision on $\underline{x}$ and $\tau_s$ and the discount factor $\alpha$: 
% \begin{equation}\label{eq:opt-in}
%     \begin{aligned}
%         U^\textrm{in}(\underline{x},\tau_s) &= \mathbb{E}_{\tau_a}\left[\alpha^{\tau_a-1}r_P\mathbf{1}_{[0,\tau_s]}(\tau_a)+\alpha^{\tau_s-1}r_Q\mathbf{1}_{(\tau_s,\infty)}(\tau_a) + \sum_{t=0}^{\tau_a\wedge \tau_s-1}\alpha^tr_t \right]\\
%         &\begin{aligned}
%             = \mathbb{E}_{\tau_a}\Bigg[\alpha^{\tau_a-1} &R\mathbf{1}_{[0,\tau_s]}(\tau_a)\ -\ C_X(x_0) \\
%             &-\sum_{t=1}^{\tau_a\wedge \tau_s-1}\alpha^t\left(C_X(x_t)-C_X(x_{t-1})\right)\Bigg] \\
%         \end{aligned}
%     \end{aligned}
% \end{equation}
\begin{equation}\label{eq:opt-in}
        U(\underline{x},\tau_s) 
        % &= \mathbb{E}_{\tau_a}\left[\sum_{t=1}^{\tau_a\wedge\tau_s} \alpha^{t-1}r_t - \sum_{t=0}^{\tau_a\wedge\tau_s-1}\alpha^tc_t\right] \nonumber\\
        = \mathbb{E}_{\tau_a}\left[\sum_{t=0}^{\tau_a\wedge\tau_s-1}\alpha^t\left(r_{t+1}-c_t\right)\right] \nonumber\\
        % &=& \mathbb{E}_{\tau_a}\Bigg[ -\ C_X(x_0)-\sum_{t=1}^{\tau_a\wedge \tau_s-1}\alpha^t\left(C_X(x_t)-C_X(x_{t-1})\right) \nonumber\\ 
            % && + \alpha^{\tau_a-1} R\mathbf{1}_{[0,\tau_s]}(\tau_a) \Bigg]. 
\end{equation}
where 
% $\mathbf{1}_{A}(x)$ is the indicator function that returns 1 when $x\in A$ and 0 otherwise, and 
$\tau_a\wedge \tau_s:=\min\{\tau_a,\tau_s\}$. Note that $r_{\tau_a}=R$ and $r_{\tau_s}=0$ as described earlier. 
%\com{So the reward at $t=1^{+}$ is not discounted, but investment decision made at $t=1^{+}$ is discounted -- the idea being the cost is spread over the period from $t=1$ to $t=2$? perhaps it's better to say decision of $x_1$ is made at time $t=1^{+}$, but the cost $C_X(x_1)=C_X(x_0)$ is incurred at time $t=2^{-}$??  This way the discounting is always consistent with what the time index $t$ is...} \resp{Yes, I agree and Fig. 1 and Eqn (2) now tries to cover this point.}
%If the vendor participates in the policy, it decides the total investment sequence $\underline{x}$ and the quitting time $\tau_s$ such that its 
The vendor's goal is to maximize its utility in Eqn \eqref{eq:opt-in}. 
%\com{So $\tau_s$ is treated as a constant then, judging by this equation?} \resp{Yes, $\tau_s$ is the decision variable of the vendor. It could be deterministic or random depending on the calculation.}
Denote the maximal utility by $U^{*}:=\sup_{\underline{x},\tau_s}U(\underline{x},\tau_s)$. 
%Eq. \eqref{eq:opt-in} suggests that the vendor's opt-in decision problem is an optimal stopping time problem.

% \com{total investments / accumulated investments and optimal stopping time problem}

\subsubsection*{The Auditor}
The auditor is modeled as a neutral party (without its own utility function). \rev{The {\em audit} is} the sequence of functions $\underline{p}(\underline{x}):=\left\{p_t(x_{t-1})\right\}_{t=1}^\infty$, where $p_t(x_{t-1})$ is \rev{the auditing {\em test} or {\em test function} executed at time $t$}, representing the probability of the product passing the audit at time $t$ given cumulative effort of $x_{t-1}$. 

% \rev{A security audit for software in practice is a complex task.  For tractability, we will model this as an estimation process, whereby the auditor predetermines a threshold $\delta$ and estimates whether the vendor's security effort exceeds it.} It follows that the estimate at stage $t$ can be represented as a \rev{random variable $Y_t:=x_{t-1}+W_t$} where $W_t\sim\mathcal{N}(0,\sigma^2_t)$. Thus the probability of passing the audit is $p_t(x_{t-1})=\mathbb{P}(Y_{t}\geq\delta)=\mathbb{P}(W_{t}\geq\delta-x_{t-1})=Q\left(\frac{\delta-x_{t-1}}{\sigma_{t}}\right)$ where $Q(z):=\mathbb{P}\left(Z\geq z|Z\sim\mathcal{N}(0,1)\right)=\frac{1}{2}\left(1-\textrm{erf}\left(\frac{z}{\sqrt{2}}\right)\right)$. 

% The audit is only {\em meaningful} or {\em informative} if it is correct more than $50\%$ of the time. In the threshold model described above, if $x_{t-1}\geq\delta$, then $p_t(\rev{x_{t-1}})\geq Q(0)=\frac{1}{2}$; if $x_{t-1}<\delta$, then $p_t(\rev{x_{t-1}})< Q(0)=\frac{1}{2}$. Thus, this threshold audit model is indeed informative for any parameterization $(\sigma_{t})_{t\geq 1}$.
\begin{assumption}\label{assump:threshold_audit}
    The audit process is static (time-invariant), given by $p_t(x)=p(x)$ for all $t$ where $p$ is an \rev{increasing and} continuous function from $\mathbb{R}_+$ to $[0,1]$.
    % The audit process is static (or time-invariant), given by $p_t(x)=p(x)$ for all $t=1,2,\dots$. % and some fixed function $p(\cdot)$.
\end{assumption}
\rev{In a static audit, the vendor under audit undergoes the same test repeatedly until it passes.}
As mentioned earlier, for the most part of the paper (Sections \ref{sec:opt-in} and \ref{sec:audit-quality}), the above static audit assumption is used. This assumption is relaxed in Section \ref{sec:dynamic} where we investigate a dynamic auditing method in the context of a special, linear test function $p(\cdot)$ (the increasing assumption continues to be needed).

\subsection{Preliminaries}\label{sec:method}
In light of Fig. \ref{fig:timing}, the vendor's decision process can be reformulated as a discounted-reward Markov decision process (MDP) as follows. 
%\com{I continue to feel we should be able to define the problem without $f$ being part of the state: we can simply say here is the definition of the state provided the vendor has not quit; and if it has, then any definition will do as it really doesn't matter.}\resp{Yes, and I think in this way we can also formulate the entire process, not just the continuing one, with a single indicator as well.}
Let $e_t\in\{0,1\}$ denote the {\em continuation state} of the process: $e_t=1$ if the process has terminated by time $t$ (inclusive), and $e_t=0$ if the process  proceeds into $t^{+}$. 
Define states $z_t:=(e_t,x_{t-1})\in\mathcal{Z}:=\{0,1\}\times\mathbb{R}_+$ for $t=1,2,\dots$ where $x_{t-1}$ is the cumulative investment over the first $t-1$ steps.  
% Define states $z_t:=(f_t, s_t, x_{t-1})\in\mathcal{Z}:=\{0,1\}^2\times\mathbb{R}_+$ for $t=1,2,\dots$ where $s_t:=\mathbf{1}_{[0,t]}(\tau_a)$ is the realization of the auditing outcome, $x_{t-1}$ is the cumulative security investment from the previous $t-1$ stages, and $f_t$ is a binary flag indicating whether the process has terminated before time $t$. 
% The state is fully observable at the beginning of every time step $t$. 
Given the current state $z_t$, the vendor chooses an action $u_t:=(q_t,a_t)\in\mathcal{U}:=\{0,1\}\times\mathbb{R}_+$ where $q_t=1$ when the vendor decides to quit at this stage and $a_t$ represents the vendor's additional investment in case of continuation.
% If the audit fails at time $t$, i.e., $s_t=0$, the vendor chooses an action $u_t:=(q_t,a_t)\in\mathcal{U}:=\{0,1\}\times\mathbb{R}_+$ where $q_t=1$ when the vendor decides to quit at this stage and $a_t$ represents the vendor's additional investment in case of continuation. 

% \rev{Let $w_t\sim\textrm{Bernouli}(p(x_t))$ be instantiated immediately after the vendor deciding its action $u_t$. In spired by the second line of Eqn \eqref{eq:opt-in}, we formulate an alternative (random) instantaneous reward $\tilde{\rho}(z_t,u_t,w_t)$ as follows
% \begin{equation}
%     \tilde{\rho}(z_t,u_t,w_t) = \begin{cases}
%         R - C_X(x_{t-1}+a_t) + C_X(x_{t-1}) & s_t=q_t=0\textrm{ and }w_t=1 \\
%         - C_X(x_{t-1}+a_t) + C_X(x_{t-1}) & s_t = q_t = 0 \textrm{ and }w_t = 0 \\
%         0 & s_t = 1\textrm{ or } q_t=1
%     \end{cases}.
% \end{equation}
% Concisely with indicator functions, $\rho(z_t,u_t,w_t)$ can be expressed as 
% \begin{equation}\label{eq:reward-form}
%     \tilde{\rho}(z_t,u_t.w_t) = \big[R\mathbf{1}_{\{1\}}(w_t) + -C_X(x_{t-1}+a_t)+C_X(x_{t-1})\big]\mathbf{1}_{\{0\}}(s_t)\mathbf{1}_{\{0\}}(q_t)
% \end{equation}
% and the expected reward provided the state-action pair is

Define an alternative instantaneous reward function as 
\begin{equation}\label{eq:reward-form}
\begin{aligned}
    \rho(z_t,u_t) = \big[&p(x_{t-1}+a_t)R - C_X(x_{t-1}+a_t)\\ &+C_X(x_{t-1})\big]\mathbf{1}_{\{0\}}(e_t)\mathbf{1}_{\{0\}}(q_t).
\end{aligned}
\end{equation}
Intuitively, this is the expected payoff (reward minus cost) that the vendor earns at time $t$. It is zero when either the process has stopped ($e_t=0$), or the vendor decides to quit ($q_t=1$) at time $t$.
% }
% \rev{Observe that the expected instantaneous reward is time invariant.}

The state at time $t+1$ can be updated using the tuple $(z_{t},u_t)$ as follows: 
% \rev{if $s_{t-1}=1$, then }
% first determine
% \begin{equation*}
%     f_{t+1}=\begin{cases}
%         0 & \textrm{if }f_t=s_t=q_t=0 \\
%         1 & \textrm{otherwise}
%     \end{cases};
% \end{equation*}
% if $f_{t+1} = 1$, then the process has stopped at this stage, we leave the other variables unchanged, i.e., $x_{t}=x_{t-1}$ and $s_{t+1}=s_t$; if $f_{t+1}=0$, the process continues and we update 
%\com{If we eliminate $f$, we can say the process stops if either $q=1$ or $s=1$; otherwise it continues ($q=0$ and $s=0$)} 
\begin{equation*}
    \begin{aligned}
        x_{t} = x_{t-1}+a_t\;\textrm{ and }\;
        e_{t+1}=\begin{cases}
            w_t& e_{t}=0\textrm{ and }q_t=0 \\
            1 & e_{t}=1\textrm{ or } q_t=1
        \end{cases}
        % , \quad f_{t+1}=\begin{cases}
        %     0 & \textrm{if }s_t=q_t=0 \\
        %     1 & \textrm{otherwise} 
        % \end{cases}
    \end{aligned}
\end{equation*}
where $w_t\sim\textrm{Bernoulli}(p(x_{t-1}+a_t))$. This system is a valid MDP by construction.

Let $\pi:=\{u_t\}_{t=1}^\infty$ denote an arbitrary policy. 
Define the \textit{expected total discounted reward} with initial state $z$ under policy $\pi$ as 
% Suppose $x_0$ is given and then let the initial state $z_1$ be drawn from the following distribution: $f_1\equiv 0$ and $s_1\sim\textrm{Bernouli}(p(x_0))$.  
% Define the \textit{expected total discounted reward} as
\begin{equation}\label{eq:mdp-objective}
    \begin{aligned}
        V^\pi(z) 
        &= \mathbb{E}_{\tau_a}\Bigg[\sum_{t=0}^{\tau_a\wedge\tau^\pi_s-1}\alpha^{t}\rho(z_t^\pi,u_t^\pi)\Bigg|z_0^\pi=z\Bigg] \\
        &= \mathbb{E}\Bigg[\sum_{t=0}^{\infty}\alpha^{t}\rho(z_t^\pi,u_t^\pi)\Bigg|z_0^\pi=z\Bigg],
    \end{aligned}
\end{equation}
where the superscript $\pi$ emphasizes the dependence of relevant variables on the policy $\pi$. 
% More specifically, \rev{$\tau_a=\min\{t\geq0:e_t=1\}$ is the first (random) time that the continuation status $e_t$ turns from $0$ to $1$ and 
% $\tau_s^g=\min\left\{t\geq0:\ q^g_t=1\right\}$ is the first time of $q_t=1$ in the given policy $g$}. 
The second equality holds because by definition, $\rho(z_t^\pi,u_t^\pi)=0$ if the process has stopped before time $t$, i.e., $e_t=1$.
The goal of the MDP is to find the optimal policy $\pi$ that maximizes the objective in Eqn \eqref{eq:mdp-objective}. Denote the \textit{optimal reward function} as $V^*(z):=\sup_{\pi}V^\pi(z)$ and the optimal policy as $\pi^*(z)\in\argsup_\pi V^\pi(z)$. 

It suffices to limit our attention to non-terminal states because $V^*(1,x) \equiv 0$ by Eqn \eqref{eq:reward-form} and \eqref{eq:mdp-objective}. With a slight abuse of notation, we will denote $V^*(x):=V^*(0,x)$.
Additionally, without loss of generality, we will only consider stationary policies, i.e., state-dependent and time-invariant, that depend on the state $z_t$ only through $x_{t-1}$. In other words, we are only interested in functions of the form $g:\mathbb{R}_+\to\mathcal{U}$ s.t. $u_t=g(x_{t-1})$ for every $t\geq0$. 
%\com{Are we saying such a function exists, or that we are only interested in cases when it exists?} %\resp{We are not introducing existence right now. The previous wording was not precise... What this sentence actually wants to say is that we only focus on stationary policies which is equivalent to optimizing over the function space $\mathbb{R}_+\to\mathcal{U}$.}

Using the notation above, we can express the vendor's optimal utility as follows: 
% Notice that the MDP can be viewed as the subgame of the original auditing process starting at stage $t=1$.
% Thus, by the principle of subgame perfect equilibrium (optimality) and Eqn \eqref{eq:opt-in-as-reward}, we can derive the relationship between the optimal reward function and the vendor's maximum opt-in utility function
\begin{equation}\label{eq:opt-utility}
    % \begin{aligned}
    %     U^{\textrm{in},*} = &\max_{x_0\in\mathbb{R}_+}\ -C_X(x_0) + \alpha\mathbb{E}\left[V^*(z_1)\Big|x_0\right]\\
    %     =&\max_{x_0\in\mathbb{R}_+}\ -C_X(x_0) + \alpha\mathbb{E}_{s_1}\left[V^*(0,s_1,x_0)\right].
    % \end{aligned}
    U^{*}=-C_X(0) + V^*(0).
\end{equation}

% TODO: Specify conditions required

% \rev{It is possible to convert the stopping time problem in Eqn \eqref{eq:mdp-objective} to an infinite-horizon problem by assigning zero to $\rho(z_t^g,u_t^g)$ for every $t\geq\tau_a\wedge\tau_s$ and summing over $t$ from $0$ to $\infty$.} 
The methodology used to compute $V^*$ is the Bellman equation. By Theorem 2.2 in \cite{hernandez-lerma_adaptive_1989}, $V^*$ is the unique solution to the following fixed-point (Bellman) equation, 
\begin{equation}\label{eq:bellman}
    V^*(x) = \sup_{u\in\mathcal{U}}\ \rho(e=0,x,u) + \alpha\mathbb{E}\Big[V^*(z')\Big|e=0,x,u\Big].
\end{equation}
where $z'$ represents the next state.

\section{The Optimal Strategy Under Audit}\label{sec:opt-in}

We now characterize the optimal strategy of the vendor undergoing an audit. We begin by determining when it is optimal to quit and then derive its optimal investments using the Bellman equation. All missing proofs can be found in the appendix.

\subsection{Optimal Quitting Time}
% By the definition of $V^g$ in Eqn \eqref{eq:mdp-objective}, we observe $V^g(1,s,x)=0$ for all $(s,x)$, which also implies $V^*(1,s,x)=0$. Intuitively, if the process has already stopped, there is no revenue to earn or costs to bear in the future. Thus, we focus on states that has not terminated. Define $\widetilde{V}(s,x):=V^*(0,s,x)$. Then, we immediately observe $\widetilde{V}(1,x) = R/\alpha + 0 = R/\alpha$.

We expand $V^*(x)$ using the quitting decision $q_t$ as
\begin{equation}\label{eq:v-tilde}
    \begin{aligned}
    V^*(x) 
    = \max\Big\{0,\ \sup_{y\geq x}\ &-C_X(y)+C_X(x)+ \\ 
    &\ p(y)R+\alpha(1-p(y))V^*(y)\Big\},
    \end{aligned}
\end{equation}
where the first term in the \textit{max} operator is the maximum reward-to-go for quitting, i.e., $q=1$, and the second term is for continuation. Notice,
\begin{equation*}
    \begin{aligned}
        \sup_{y\geq x}\; &C(x) -C(y)+ p(y)R + \alpha (1-p(y)) V^*(y) \\
        &\geq\  p(x)R + \alpha (1-p(x)) V^*(x) \geq 0,
    \end{aligned}
\end{equation*}
where the last inequality utilized $V^*(x)\geq 0$ by Eqn \eqref{eq:v-tilde}.
% Then, 
% \begin{equation*}\begin{aligned}
%     V^*(x)&\geq\ \sup_{y\geq x}\; C(x) -C(y)+ p(y)R + \alpha (1-p(y)) V^*(y) \\
%         &\geq\  p(x)R + \alpha (1-p(x)) V^*(x),
% \end{aligned}
% \end{equation*}
% which implies 
% \begin{equation*}
%     V^*(x)\geq\frac{p(x)R}{1-\alpha+\alpha p(x)} > 0.
% \end{equation*}
This directly leads to the following lemma.

%\com{Potentially making this a lemma...}
\begin{lemma}\label{lemma:never-quit}
    The vendor never quits in an optimal strategy.
\end{lemma}

\subsection{Optimal Continuation Investment}\label{sec:continuation-investments}
Given that a vendor never quits, we can remove the \textit{max} operator in Eqn \eqref{eq:v-tilde} and express $V^*(x)$ more concisely as 
\begin{equation*}
    V^*(x) = \sup_{y\geq x}\ C(x) -C(y) + p(y)R+\alpha(1-p(y))V^*(y).
\end{equation*}
Define $W(x):= V^*(x)-C(x)$. The above equation implies
\begin{equation}\label{eq:w}
    W(x)=\sup_{y\geq x}\ -C(y)+p(y)R+\alpha(1-p(y))V^*(y).
\end{equation}
The optimal additional investment $a^*$ given the cumulative investments $x$ thus satisfies
\begin{equation}\label{eq:optimal-action-w}
    x+a^* \in \argsup_{y\geq x}\  -C(y) + p(y)R +\alpha(1-p(y))V^*(y).
\end{equation}
There exists an optimal additional investment if the \textit{arg\;sup} in the second term yields a non-empty set.

\begin{lemma}\label{lemma:mono-w}
    $W$ is decreasing in $x$.
\end{lemma}
\begin{proof}
    Define the set $$\mathcal{B}_x:=\left\{-C_X(y) + p(y)R + \alpha(1-p(y))V^*(y):\ y\geq x\right\}.$$ Notice that $W(x)=\sup\mathcal{B}_x$ where we allow $W(x)=\infty$ if $\mathcal{B}_x$ is unbounded. For any $x_1,x_2\geq0$ and $x_1\leq x_2$, it follows $\mathcal{B}_{x_1}\supseteq de \mathcal{B}_{x_2}$, which implies $W(x_1)\geq W(x_2)$.
\end{proof}

The monotonicity of $W$ has a very interesting implication on the vendor's behavior. Suppose the cumulative investments up to $t$ is $x_{t-1}$ and the vendor chooses an $a^{*}_t$ optimally according to Eqn \eqref{eq:optimal-action-w}, resulting in a new cumulative investment $x_t=x_{t-1}+a_t^*$. If the vendor fails the test at this level, its optimal additional investment now becomes zero since the maximum in Eqn \eqref{eq:w} is already obtained with $x_t$ over $[x_{t},\infty)$ by the monotonicity of $W$.
% and \rev{optimality of $x_{t}^*$}
Therefore, one of the vendor's optimal strategies given any cumulative  investment (sunk cost) is to immediately invest at the optimal additional level and wait indefinitely until it passes the audit.

\begin{lemma}\label{lemma:explicit-w}    
    The function $W$ can be expressed as
\begin{eqnarray}\label{eq:opt-W-G}       
W(x)=\sup_{y\geq x}\ G(y),\quad\forall x\geq 0,    \end{eqnarray}
    where    \begin{equation}\label{eq:G-def}
        G(y):= -C_X(y)+\frac{ p(y)R}{1-\alpha+\alpha p(y)}.
    \end{equation} 
     Suppose the optimal additional investment $a^*(x)$ exists for any cumulative investment $x$.
     % if and only if the set $\mathcal{G}_0:=\argsup_{x\geq0}G(x)$ is non-empty and is given by
     Then, the set $\mathcal{G}_x:=\argsup_{y\geq x}G(y)$ is non-empty for every $x$ and the optimal additional investment satisfies
\begin{equation}\label{eq:opt-invst-G}
        x+a^*(x)\in\mathcal{G}_x.
\end{equation}
\end{lemma}

\begin{proof}
    According to Eqn \eqref{eq:w}, $W$ can be expressed as
    \begin{multline*}
            W(x)=\ \sup_{y\geq x}\ p(y)R - (1-\alpha + \alpha p(y)) C_x(y) \\ + \alpha (1-p(y)) W(y) \\
            \quad\quad\geq   p(x)R - (1-\alpha + \alpha p(x))C_x(x) + \alpha (1-p(x)) W(x), \\
    \end{multline*}
    which implies
    \begin{equation*}
        W(x) \geq - C_X(x) + \frac{ p(x)R}{1-\alpha + \alpha p(x)},
    \end{equation*}
    as $1-\alpha + \alpha p(x) > 0$ for every $x>0$.
    For any $x\geq 0$ and $y\geq x$, by Lemma \ref{lemma:mono-w}, we have 
    \begin{equation}\label{eq:proof-geq}
        \begin{aligned}
            &W(x)\geq W(y) \geq - C_X(y) + \frac{ p(y)R}{1-\alpha + \alpha p(y)} \\
            \implies &W(x)\geq\sup_{y\geq x}\ -C_X(y)+\frac{ p(y)R}{1-\alpha+\alpha p(y)}.
        \end{aligned}
    \end{equation}

    To show the reverse direction, define the function over which $W(x)$ takes supremum as
    \begin{equation*}
        g(y) := - (1-\alpha + \alpha p(y))C_X(y) +  p(y)R + \alpha (1-p(y)) W(y).
    \end{equation*}
    
    By assumption, there exists $\overline{y}_x:= x + a^*(x)$ such that $g(\overline{y}_x)=W(x)$ (the $sup$ is attainable). Then, by monotonicity of $W$, we have $g(\overline{y}_x)=W(x)\geq W(\overline{y}_x)\geq g(\overline{y}_x)$ $\implies$ $W(x)=W(\overline{y}_x)=g(\overline{y}_x)$. The latter inequality can be reorganized into
    \begin{equation*}
        W(\overline{y}_x) = -C_X(\overline{y}_x) + \frac{ p(\overline{y}_x)R}{1-\alpha + \alpha p(\overline{y}_x)}=G(\overline{y}_x),
    \end{equation*}
    where the second equality comes from the definition of $G(\cdot)$, which further implies $W(x)=G(\overline{y}_x)$. It then follows from Eqn \eqref{eq:proof-geq} that $\overline{y}_x\in\mathcal{G}_x$, implying the latter is non-empty.
    Since $\overline{y}_x\geq x$, we have
    \begin{equation}\label{eq:reverse}
    \begin{aligned}
        W(x)&=G(\overline{y}_x)\leq\sup_{y\geq x}\ G(y)\\&=\sup_{y\geq x}\ -C_X(y)+\frac{ p(y)R}{1-\alpha+\alpha p(y)}.
    \end{aligned}   
    \end{equation}
    Together with Eqn \eqref{eq:proof-geq}, we conclude $W(x)=\sup_{y\geq x}G(y)$.
\end{proof} 

Comparing the maximum utility value in Eqn \eqref{eq:opt-utility} and the definition of $W$ in Eqn \eqref{eq:w}, we see that $U^{*}=W(0)$. Therefore, we can directly calculate the optimal (sequential) investments by evaluating the function $G(x)$. 

% However, extra care should be taken when dealing with multiple solutions to Eqn \ref{eq:opt-invst-G}.

% Let's plug in $p(x)$ using the static threshold audit example described in Section \ref{sec:problem-setup}. We obtain
% \begin{equation*}
%     G(x) = -\exp(-bx)-cx + \frac{Q\left(\frac{\delta-x}{\sigma}\right)R}{1-\alpha + \alpha Q\left(\frac{\delta-x}{\sigma}\right)},
% \end{equation*}
% and its derivative
% \begin{equation*}
%     G'(x)=b\exp(-bx) - c + \frac{1-\alpha}{\sigma} \frac{f\left(\frac{\delta - x}{\sigma}\right)}{(1-\alpha+\alpha p(x))^2}R,
% \end{equation*}
% where $f$ is the probability density function of the standard normal distribution.

% In summary,
\begin{theorem}\label{thm:opt-in-optimal}
    Suppose the optimal additional investment exists for any cumulative investment $x$. The vendor's optimal strategy under audit has to satisfy the following properties:
    \begin{enumerate}
        \item[(1)] the vendor will never quit; 
        \item[(2)] it is given by any non-decreasing sequence of cumulative investments $\{x_t\}_{t\geq 0}$, where  $x_t\in\mathcal{G}_0$; %any such sequence is also guaranteed to be optimal;}
%        \com{Okay I will be conservative here and not over reach, though I don't think it matters because the system is history independent..}
        %\resp{Actually, the "only if" is a little tricky because any optimal "Markov" policy should satisfy (2), but if we consider historically dependent policies then there could exist other equally optimal ones. The MDP theory shows Markov policies are sufficient but not the only ones. But we did restrict ourselves to Markov policies right?} %\com{we know any such sequence is optimal; is it also the case that any optimal sequence must also be such a sequence? That is, is this if and only if?}\resp{Actually, the "only if" is a little tricky because any optimal Markov policy should satisfy (2), but if we consider historically dependent policies then there could exist equally optimal ones.}
        %  initially invests $x_0^*$ and wait with no additional investments until the audit passes, where $x_0^*$ is given by 
        % \begin{equation*}
        %     x_0^* \in\argmax_{x_0\geq 0}\ G(x_0),
        % \end{equation*}
        \item[(3)] the optimal utility is given by $\max_{x\geq0}G(x)$.
    \end{enumerate}
\end{theorem}
% \com{example on explanation of (2), x1,x2,x3 example}
% Notice that there might be other strategies in which investments are not concentrated in the initial stage if the set $\argmax_{x_0\geq0}G(x_0)$ contains more than one elements, or $G(\cdot)$ has multiple global maximum over $[0,\infty)$. In that case, the vendor is indifferent between exerting zero additional investments and raising the investment to a different global maximum. We break ties by assuming the vendor will always choose the largest investment and exert zero additional efforts in the subsequent auditing process.  

Theorem \ref{thm:opt-in-optimal}-(2) says that the optimal strategy is in general non-unique, but it must fall into two broad categories. The first type of optimal strategy is such that the vendor invests any amount $x\in \mathcal{G}_0$ at stage $0$ followed by nothing else in subsequent stages, essentially waiting for the audit to return a positive outcome (which is guaranteed to occur with high probability given our assumptions on the audit process). 
%\com{yes??}\resp{Yes, because the probability of not passing at all is $(1-p(x))^\infty\approx 0$.} 
This class of strategies can be referred to as the ``patient'' type, deciding on a total expenditure and then waiting it out.  In particular, those that invest in the smallest amount (the smallest $x$ in $\mathcal{G}_0$) are foregoing revenue (due to the long expected time it takes to pass the audit, the eventual revenue would be severely discounted) in exchange for a small initial investment. 

The other class of optimal strategies involves investing at two or more different times, each time reaching some cumulative amount $x\in \mathcal{G}_0$. When these investments are made is arbitrary, provided the first occurs at time $0$. This type of strategy is more impatient or opportunistic: they invest some small amount initially, hoping to pass the audit on good luck; when that doesn't happen for some time, they decide to up the game and invest more hoping to pass the audit this time, and so on. 

It is important to emphasize that both types of strategies yield the same utility under our model; they essentially reflect different tradeoffs between willingness to invest vs. willingness to wait for return on investment. 

Those who invest the largest amount $x\in\mathcal{G}_0$ at time $0$ necessarily belong to the first type, as there is no more feasible action left given the non-decreasing nature of the sequence. These are the ``ideal'' or most desirable vendors from a public interest or social welfare perspective -- they invest the maximum amount in one go, resulting in the highest quality product. In the next section, we will discuss which configurations of the auditing process can help lead to this type of strategy. 

The following example shows more concretely the property of the optimal strategies given in Theorem \ref{thm:opt-in-optimal}-(2) and discussed above.

\begin{figure}[!htbp]
    \centering
    \includegraphics[width=.8\linewidth]{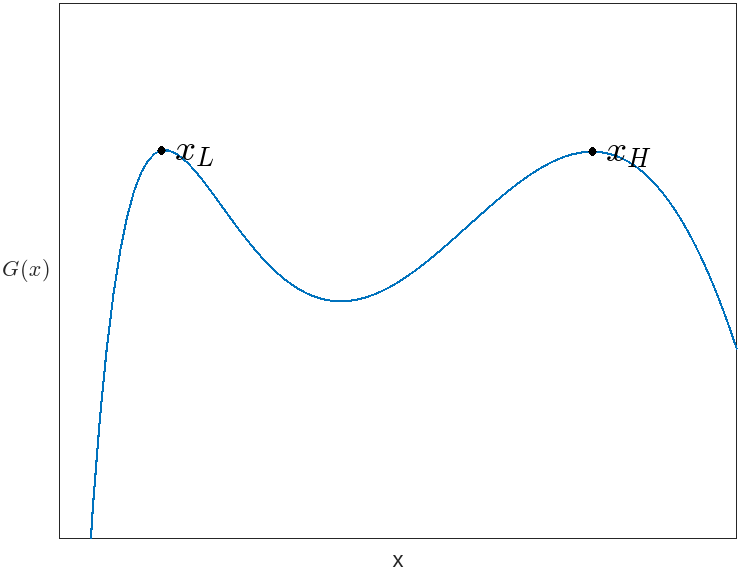}
    \caption{Shape of $G(x)$ with two local maxima $x_L$ and $x_H$ with $x_L<x_H$; in this example both attain the same global maximum.}
    \label{fig:two-peaks}
\end{figure}

\begin{example} (Property of the Optimal Sequence of Cumulative Investments)
    Suppose $G(x)$ contains two local maximizers $x_L$ and $x_H$ with $x_L<x_H$ as illustrated in Fig. \ref{fig:two-peaks}. If $G(x_L)\neq G(x_H)$, then there is a single global maximum, and the optimal investment strategy is unique: investing at the global maximum level at time 0 and nothing else thereafter.
    
    If $G(x_L)=G(x_H)$, then the set $\mathcal{G}_0$ in Theorem \ref{thm:opt-in-optimal} contains exactly 2 values $x_L$ and ${x}_H$. By Theorem \ref{thm:opt-in-optimal}-(2), every optimal investment sequence should start with $x_0\in\{{x}_L,{x}_H\}$. In other words, the initial action $a_0^*\in\{{x}_L,{x}_H\}$. 
    %as $x_0=x_{-1}+a_0^*=a_0^*$ by our definition.   

    If an optimal strategy starts with $x_0={x}_H$, then all subsequent cumulative investments remain at ${x}_H$, i.e., no additional investment in the future. In this case, the optimal cumulative investment sequence and the optimal action sequence are, respectively,
    \begin{equation*}
        \{x_t\}_{t\geq0}=\{{x}_H,{x}_H,\dots\}\;\textrm{ and }\;\{a_t^*\}_{t\geq0} = \{{x}_H,0,0,\dots\}.
    \end{equation*}
    As ${x}_H$ is the largest element in $\mathcal{G}_0$, $\{{x}_H,{x}_H,{x}_H,\dots\}$ is the unique non-decreasing sequence of cumulative investments in $\mathcal{G}_0$ given $x_0={x}_H$. This strategy minimizes the vendor's expected time for passing the audit. 
    %We call this type of vendors as {\em conservative} for choosing higher investments at early stages with higher likelihood to secure an early pass.
    
    If an optimal strategy starts with $x_0={x}_L$, then this can lead to either type of optimal sequence.  Under the first type, the vendor invests nothing more beyond the initial amount, i.e.,
    \begin{equation*}
        \{x_t\}_{t\geq0}=\{{x}_L,{x}_L,\dots\}\;\textrm{ and }\;\{a_t^*\}_{t\geq0} = \{{x}_L,0,0,\dots\}. 
    \end{equation*}
    Under the second type, the vendor invests an additional ${x}_H-{x}_L$ at some arbitrary future time $t\geq 1$, i.e., 
    \begin{align*}
        &\{x_t\}_{t\geq0}=\{{x}_L,x_L,\dots,{x}_L,{x}_H,\ \ {x}_H,\dots\}, \\
        &\{a_t^*\}_{t\geq0} = \{{x}_L,0,\dots,0,{x}_H-{x}_L,0,\dots\}. 
    \end{align*}
%where $x_T$ is the first $\tilde{x}_3$ and $a_T=\tilde{x}_3-\tilde{x}_2$. 
As mentioned earlier, the exact time at which the additional investment ${x}_H-{x}_L$ is made has no impact on the strategy's optimality. %This type of vendors are {\em opportunistic} for choosing smaller investments at early stages, hoping for a lucky pass.
% value of the time $T$ has no influence on the optimality. The vendor can switch to a higher cumulative investment at any time without losing its utility. 

% Similarly, if the optimal strategy starts with $x_0=\tilde{x}_1$, then the vendor could either invest nothing more, or increase its cumulative investment to $\tilde{x}_2$ (or $\tilde{x}_3$) at some arbitrary time in the future but nothing more, or increase to $\tilde{x}_2$ at some time, followed by another increase to $\tilde{x}_3$ some time later. 
%, then the remaining optimal sequence $\{x_t\}_{t\geq T}$ is the same as the optimal cumulative investment sequence in case of $x_0=\tilde{x}_2$. If it decides to directly increase to $\tilde{x}_3$ at some time $T$ in the future, then it should invest nothing additional after time $T$.

\end{example}

\vspace{-10pt}

% Then, we obtain
% \begin{equation*}
%     G(x) = -\exp(-bx)-cx + \frac{Q\left(\frac{\delta-x}{\sigma}\right)R}{1-\alpha + \alpha Q\left(\frac{\delta-x}{\sigma}\right)},
% \end{equation*}
% and its derivative
% \begin{equation*}
%     G'(x)=b\exp(-bx) - c + \frac{1-\alpha}{\sigma} \frac{f\left(\frac{\delta - x}{\sigma}\right)}{\left[1- \alpha F\left(\frac{\delta-x}{\sigma}\right)\right]^2}R,
% \end{equation*}
% where $f$ is the probability density function (PDF) and $F$ is the cumulative distribution function (CDF) of the standard normal distribution. One can show (see Section \ref{}) that the last term in the above equation increases in $[0,x']$ for some $x'\geq0$ and then decreases in $(x',\infty)$.

\subsection{Equivalence to Liability Insurance}

There is an interesting interpretation of the static auditing mechanism from a liability insurance perspective.  By Theorem \ref{thm:opt-in-optimal}, we can re-write the utility function for opt-in vendors as follows:
\begin{eqnarray}
        U^\textrm{in}(x) &=& G(x) = -C_X(x) + \frac{p(x)R}{1-\alpha + \alpha p(x)} \nonumber \\ 
        &=& R - C_X(x) - \frac{(1-\alpha)(1-p(x))R}{1-\alpha + \alpha p(x)} \nonumber \\
        &=:& R- C_X(x) - C_A(x; p) \label{eq:alt_liability}~, 
\end{eqnarray}
with the optimal opt-in strategy obtained by maximizing $U^\textrm{in}(x)$ over $[0,\infty)$. 
Comparing the above expression to the opt-out utility function given in Eq. \eqref{eq:opt-out}, we see the two only differ in their last terms: liability loss $C_L(x)$ in Eq. \eqref{eq:opt-out} and $C_A(x;p)$ in Eq. \eqref{eq:alt_liability}. 

\rev{
This comparison provides an alternative interpretation of the optimal opt-in strategy. It suggests that the audit mechanism is equivalent to a ``waiver-for-fee'' mechanism, i.e., offering the vendor complete liability waiver in exchange for a one-time fee of $C_A(x;p)$.  This is nothing but an insurance policy with premium discrimination.  Viewed through the insurance lens, $C_A(x;p)$ is functionally equivalent to the premium charged by the insurance provider; it is assessed not only based on the security effort of the vendor (i.e., investment $x$), but also on the market value of the product (i.e., potential revenue $R$). The insurance provider may or may not perform an audit as long as it has a way of determining $p(x)$. Note, however, that this comparison to insurance merely serves as an alternative interpretation of the audit mechanism, but does not address whether such an insurance provider would indeed exist and makes profit. This is a crucial difference between a profit-maximizing insurer and a profit-neutral auditor modeled in this paper; a more comprehensive comparison will be an interesting direction of future study.
} 

\subsection{Risk Perception under the Audit Mechanism}\label{sec:risk}
The shape of this function $C_A(x;p)$ reveals quite a few interesting properties. 
Firstly, this function is decreasing in $x$ with diminishing margins, similar to the liability loss function $C_L(\gamma,x)$. When the vendor decides to opt in, this waiver cost essentially replaces the liability loss and thus represents the ``risk'' now perceived by the vendor. 

\begin{figure}[!htbp]
    \centering
    \includegraphics[width=0.8\linewidth]{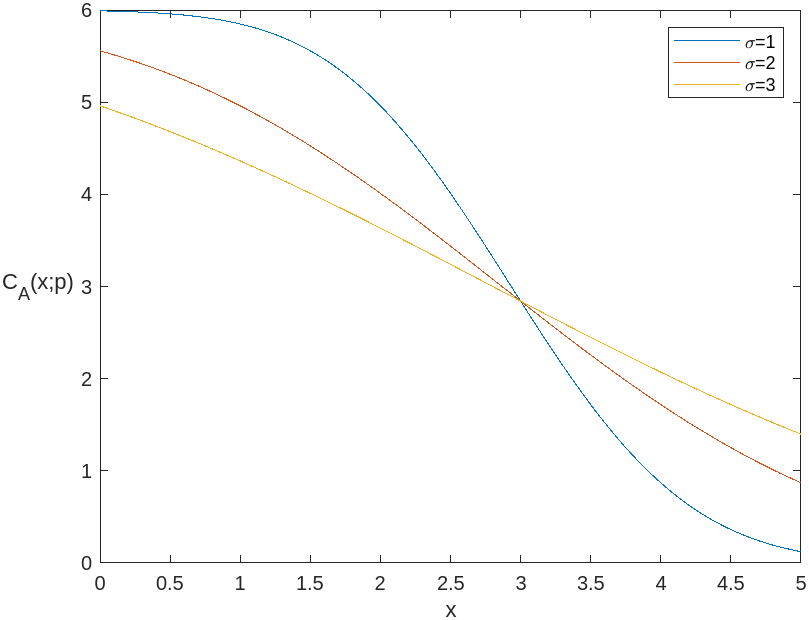}
    \caption{Auditing cost function $C_A(x;p)$ with threshold audit example evaluated at different noise levels $\sigma$, taking the threshold as fixed at $\delta=3$.}
    \label{fig:ca}
\end{figure}

A few examples of this function with $p(x)$ being the threshold form defined in Section \ref{sec:problem-setup} are depicted in Fig. \ref{fig:ca}. 
We observe that $C_A(x;p)$ is first concave and then convex as $x$ increases. This suggests that under the audit mechanism, the vendor's risk attitude transitions from risk seeking to risk aversion; 
%\com{need to double check which is which...} \resp{I feel like the concave part, (smaller $x$ part) is risk seeking and the convex part is (larger $x$ part) is risk averse. It is because Fig. 3 is defined in terms of the loss rather than utility. So the risk averse part is the concave part in $-C_A(x;p)$ and thus convex part in $C_A(x;p)$. } \com{I agree; this is what I thought -- we need to rephrase this part...}
%, i.e., $C_L(\gamma,x)$, into a combination of both risk-seeking and risk-aversion, i.e., $C_A(x;p)$. 
the former dominates at lower investment levels, while the latter at higher investment levels.
% \rev{The point of change occurs at $x=\delta$. Since the audit is informative, the probability of pass is less than $0.5$ when $x<\delta$,  }
%\com{I'm inclined to suggest we keep the figure but end the discussion here, and remove the next two paragraphs... I feel more effort is needed to clarify what we mean here.  Also, I know think the optimal strategy has been sufficiently explained in the previous section so perhaps we don't need to beat up on this more?} 
Notice that $C_A(x;p)$ is presented as loss so that the risk-averse (resp. risk-seeking) region corresponds to the concave (resp. convex) region of $-C_A(x;p)$. This is in contrast to $-C_L(x)$ which is purely concave, indicating risk aversion under any security investment. 

% The interpretation is that a vendor who is currently in a low investment region (i.e., it has so far made little effort) tends to see the probability of passing the audit as minimal. Thus, it may take a risk by investing even less, betting on the probability of a favorable outcome. \com{This is risk aversion in our case??} 
% On the other hand, the vendor in higher investment region is more determined to secure a successful audit and thus tends to invest even more assertively. 

% This pattern leads to the consequence of the vendor's investment level locating in extreme ends, either very low or very high, corresponding respectively to the attitudes of opportunism and conservatism described in the previous section. This also gives rise to a secondary solution to the opt-in utility-maximization problem above.

\section{Incentivizing Audit Participation}\label{sec:audit-quality} 

The previous section focuses on the vendor's optimal strategy to pass the audit.  We now turn to the question of how the audit mechanism should be designed so as to encourage participation by the vendor if it is voluntary. This is the {\em voluntary participation} problem extensively studied in the mechanism design literature. We will continue to assume that once the vendor chooses to go through the audit, it cannot release the product until it passes the audit. In return, passing the audit relieves the (vendor of the) product from potential future liabilities \rev{either in full or partially (i.e., the vendor obtains a full or partial liability waiver)}. 
This is the scenario we will model below. 
 However, one might equivalently view the audit as a certification process, whereby passing the audit earns the vendor a stamp of approval that carries certain reputational or pricing benefits.

On the other hand, if the vendor decides to opt-out, then it will bear the cost of any potential liability on its product in the case of an adverse event (or equivalently, it will not obtain the reputational benefit of certification). Thus the availability of such an audit service may be viewed as a type of mechanism aimed at incentivizing a vendor to increase its effort and subject its product to audit.  

\rev{It is important to note that we do not make a distinction between a full or a partial waiver as it is inconsequential to our analysis; our model simply affords those who have passed the audit a certain reward, which is a function of the auditor's own effort as detailed below.}
The central questions we seek to address are: Under what conditions would a vendor voluntarily participate in such a mechanism, and can the audit induce better/higher effort from the vendor?

If the vendor opts out of the auditing mechanism, its optimal action is to choose some $x\in\mathbb{R}_+$ at $t=0^{+}$ that maximizes the following {\em expected} opt-out utility, which is attained at time $t=1$ when the product is put on the market: 
\begin{equation}\label{eq:opt-out}
    U^{\textrm{out}}(x):=R-C_X(x)-C_L(x),
\end{equation}
where $C_L(x)$ represents the potential liability cost/loss \rev{(or the loss differential between opting out of the audit and passing the audit, thereby suffering lower loss due to a full or partial waiver; for simplicity, this will be simply referred to as the loss for the remainder of this section)}. While the reward and development cost terms are assumed deterministic, losses are random in general. Thus, $C_L(x)$  denotes the expected loss perceived by the vendor, with potential risk aversion built in. We discuss this next.

\subsubsection*{Specific Functions Used in the Analysis} 
We will denote the actual monetary liability loss (in USD) by the random variable $Z(x)$, assumed to follow a normal distribution with mean $\mu_Z(x)$ and standard deviation $\sigma_Z(x)$. We will assume $\mu_Z(x)$ and $\sigma_Z(x)$ are both positive and decreasing in $x$, i.e., higher effort reduces the expected loss and the uncertainty in the loss. 
% The density function of $Z(x)$ is 
%\com{Is there a reason to introduce a new notation $Z_x$ rather than using what you have already defined, $C_L(x)$? -- Okay, I see that you are using this as the expected, or risk-averse version of the cost..} 
% \begin{equation*}
%     f(z;x)=\frac{1}{\sigma_Z(x)}\frac{\varphi\left(\frac{x-\mu_Z(x)}{\sigma_Z(x)}\right)}{1-\Phi\left(-\frac{\mu_Z(x)}{\sigma_Z(x)}\right)},
% \end{equation*} 
% where $\varphi(\cdot)$ and $\Phi(\cdot)$ are respectively the probability density and cumulative distribution functions of the standard normal distribution. 
% \cite{dowson_maximum-entropy_1973} shows that truncated normal is the maximum entropy distribution on $[0,\infty)$ provided the knowledge of first and second moments subjected to the condition $\mu_2<2\mu_1^2$ where $\mu_k$ is the $k$-th moment. 
% We argue that representing the liability loss with known mean and variance with truncated normal distributions is a suitable choice by the principle of maximum entropy. It states that the maximum entropy distribution is the least informative, therefore the best, distribution that represents the current system with given prior knowledge of moments \cite{PhysRev.106.620}. 

To capture the vendor's risk aversion, we will model the liability cost that enters into the vendor's utility function as 
%Liability loss should be transformed through a risk aversion function before entering the vendor's utility function. Define 
$C_L(x) := %C_L(\gamma,x) = 
\mathbb{E}\exp(\gamma Z(x))$, where $\gamma>0$ represents the vendor's risk attitude. 
% An interesting observation on the Gaussian assumption is that the {\em perceived} risk after the transformation through risk aversion, i.e., $\exp(\gamma Z_x)$, actually follows a log-normal distribution, which belongs to the heavy-tailed distribution family\footnote{It is worth noticing that the cyber loss itself, i.e., $Z_x$, is usually directly seen/modeled as heavy-tailed.  \rev{Mathematically, our assumption of Normally distributed loss filtered through a risk-averse utility function achieves a similar effect.}
%We simplified the assumption in this work by assuming the truncated normal distribution. It is not common that the risk-averse-transformation of arbitrarily distributed loss follows a heavy-tail distribution and this distribution is controlled by the risk-aversion parameter $\gamma$.

By the property of normal distribution, we can write $C_L(\gamma,x)$ as 
\begin{equation}\label{eq:liability-expression}
\begin{aligned}
    C_L(\gamma,x)&=\exp\left(\gamma \mu_Z(x) + \frac{1}{2}\gamma^2\sigma_Z^2(x)\right).
\end{aligned}
\end{equation}

We model the test function $p(\cdot)$ as an estimation process, whereby the auditor predetermines a threshold $\delta$ and estimates whether the vendor's effort exceeds it. It follows that the estimate, given the vendor's effort $x$, can be represented as a random variable $Y:=x+W$ where $W\sim\mathcal{N}(0,\sigma^2)$ is random noise with normal distribution. \rev{
The presence of noise highlights the fact that no test can be perfect. The normal assumption models various unknown sources of uncertainty; its variance represents the {\em accuracy} or {\em quality} of the test: a more accurate test has higher certainty.}
The probability of passing this specific test is $p(x)=\mathbb{P}(Y\geq\delta)=\mathbb{P}(W\geq\delta-x)=1-\Phi\left((\delta-x)/\sigma\right)$, where $\Phi(\cdot)$ is the CDF of the standard normal distribution. 

% $=Q\left(\frac{\delta-x}{\sigma}\right)$ where $Q(z):=\mathbb{P}\left(Z\geq z|Z\sim\mathcal{N}(0,1)\right)=\frac{1}{2}\left(1-\textrm{erf}\left(\frac{z}{\sqrt{2}}\right)\right)$. 

The test is only {\em meaningful} or {\em informative} if it is correct more than $50\%$ of the time. The above threshold model is indeed meaningful: if $x\geq\delta$, then $p(x)\geq 1-\Phi(0)=\frac{1}{2}$; if $x<\delta$, then $p(x)< 1-\Phi(0)=\frac{1}{2}$. 
% Thus, this threshold audit model is indeed informative for any parameterization $(\sigma_{t})_{t\geq 1}$.
% \begin{assumption}\label{assump:threshold_audit}
%     The audit process is static (or time-invariant), given by $p_t(x)=p(x)$ for all $t=1,2,\dots$ and some fixed function $p(\cdot)$.
% \end{assumption}
% Applying the above assumption in the threshold audit yields $\sigma_t=\sigma$ for all $t=1,2,\dots$ and some fixed $\sigma>0$.

For the cost of investment, we adopt a linear form where the marginal cost of investment for the vendor is constant, i.e., $C_X(x):=c\cdot x$ for some $c>0$. 
We note that the specific functional form of this marginal cost is not critical to the subsequent analysis, as our results hold without the linearity or even the monotonicity of this cost function\cite{huang_incentivizing_2024}.

While we do not model the auditor as a strategic agent, the vendor's strategy, and moreover, its choice of participation, is indeed influenced by the test threshold $\delta$ and test noise $\sigma$. Below, we first examine how these test parameters impact the vendor's strategy when it opts in and then show how they impact the vendor's decision to opt in vs. stay out. For clarity, we will refer to terms introduced prior to Section \ref{sec:audit-quality} with the prefix ``opt-in'', such as ``opt-in utility'' and ``opt-in strategy'' to distinguish them from Problem \eqref{eq:opt-out}. 

\subsection{On the Vendor's Optimal Opt-In Strategy}

%\zy{I adjusted the figures and this section a little as we now have different liability and cost functions.}
%\com{Cost is different but liability is the same, no?}\resp{Not quite the same because the liability now is not truncated.}

Results and discussion in Section \ref{sec:continuation-investments} suggest there can potentially be many optimal strategies for an opt-in vendor, some starting at very low investment levels depending on the solution set to $G(x)$.   While these are equally optimal by the definition of our model, the auditor may favor earlier and higher investments. Below we show that different choices of $\delta$ and $\sigma$ can reshape $G(x)$ so as to induce more desired opt-in strategies.  

\begin{figure*}[!t]
    \centering
    \subfloat[$G(x)$ with varying $\delta$ under fixed $\sigma=1$. The stars mark the (unique) global maximum of $G(x)$.]{\includegraphics[width=0.35\textwidth]{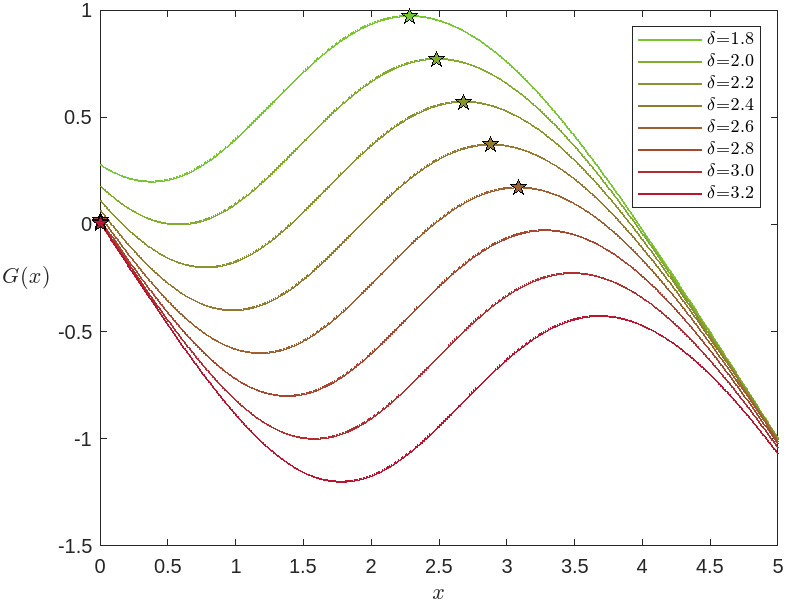}%
    \label{fig:varying-delta}}
    \hfil
    \subfloat[$G(x)$ with varying $\sigma$ under fixed $\delta=1$. The stars mark the (unique) global maximum of $G(x)$.]{\includegraphics[width=0.35\textwidth]{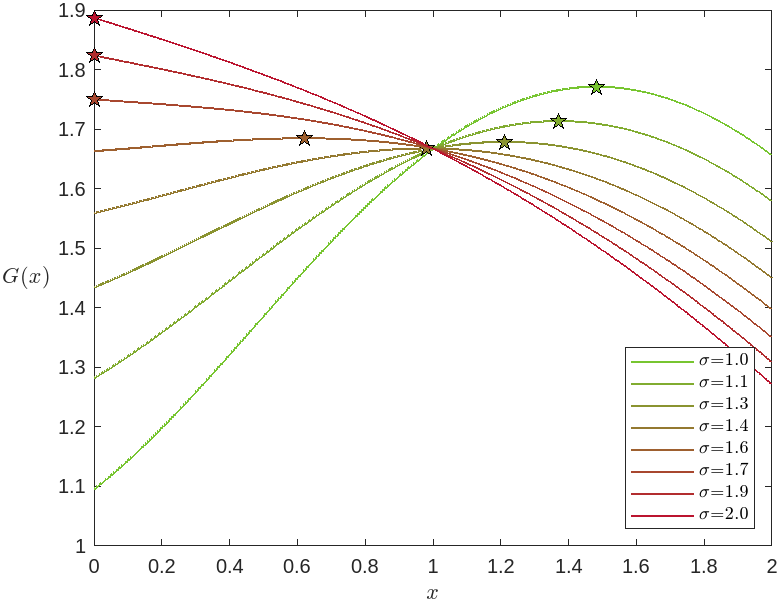}%
    \label{fig:varying-sigma}}
    \caption{Influence of auditing setup on vendor's utilities.}
    \label{fig:varying-audit}
\end{figure*}

Fig. \ref{fig:varying-audit} depict the shape of $G(x)$ under different values of $\delta$ and $\sigma$ respectively, while keeping the other fixed\footnote{Other parameters are $c=1$, $\alpha=0.5$, $R=4$.}. The global maximum solutions of each curve correspond to the optimal opt-in efforts in that specific parameter setup, marked by solid stars in the figures.  The main observations are summarized as follows: 
%\com{The wording here is generally quite confusing: I wrote what I think you are saying but please verify. I also commented out a lot of the text: we should aim for clarity and conciseness here, not volume...} \resp{Yes, that is totally correct... }

\begin{enumerate}
%\item As seen from Fig. \ref{fig:varying-delta}, a low (easier) audit threshold $\delta$ is more likely to result in a single maximum solution to $G(x)$; whereas a high (more difficult) threshold $\delta$ is more likely to result in multiple maximum. 

\item From Fig. \ref{fig:varying-delta}, we see when two local maximizers exist in $G(x)$, a high threshold (more difficult test) causes the low solution $x=0$ to dominate (it becomes the global maxima), whereas a low threshold (an easier test) leads to the high solution dominating (it becomes the global maxima). 
 %Fig. \ref{fig:varying-delta}. Notice the threshold boundary (dashed black horizontal line) in Fig. \ref{fig:varying-delta}. 
 % In the region where the audit threshold is small ($\delta \leq 2.5$) and $x_H$ induces higher utility than $x_L$; in the region below the boundary, the audit threshold is large and $x_L$ results in higher utility.) 
 Also, a lower threshold always results in higher utility for the vendor, regardless of the effort.
This suggests that a high threshold $\delta$ can encourage low effort as an optimal strategy.  This seems counter-intuitive; the reason is that a difficult test poses a risk of failing the test even at decent effort levels, so the vendor invests less and instead relies on waiting for a positive test outcome to materialize by chance. 
%\com{This is rather counter-intuitive, no?  We are saying making these audit hard to pass is not a good thing, yes?} 
A low threshold, on the other hand, reduces the need to gamble on the outcome of the test and encourages the vendor to invest at the optimal (high) level from the start, aimed at ensuring a speedy pass. 

\item From Fig. \ref{fig:varying-sigma}, we see that the shape of the $G(x)$ function is even more sensitive to the test noise: a low noise (greenish curves) drives the high solution to become the global maximum (thus high investment as an optimal strategy), while a high noise (reddish curves) drives the low solution to be the global maximum (thus low investment as an optimal strategy). 
\end{enumerate} 

In short, the above observations suggest that an accurate (low noise) but not overly strict/difficult test/audit (so it is possible to pass) is the best choice: it minimizes opportunistic behavior and reliance on chance and encourages higher levels of effort early on in the process. 

% \begin{figure*}[!htbp]
%     \centering
%     \includegraphics[width=\textwidth]{figure/varying-delta-2.png}
%     \caption{$G(x)$ with varying $\delta$ under fixed $\sigma=1$. The stars mark the (unique) global maximum of $G(x)$.}
%     \label{fig:varying-delta}
% \end{figure*}

% \begin{figure*}[!htbp]
%     \centering
%     \includegraphics[width=\textwidth]{figure/varying-sigma-3.png}
%     \caption{$G(x)$ with varying $\sigma$ under fixed $\delta=1$. The stars mark the (unique) global maximum of $G(x)$.}
%     \label{fig:varying-sigma}
% \end{figure*}

\subsection{On the Vendor's Choice of Participation}
Clearly, the vendor only has the incentive to participate in the audit mechanism if $U^{*}\geq U^{\textrm{out},*}$ (voluntary participation), assuming ties are broken in favor of participation. 
% \rev{
% In light of Theorem \ref{thm:opt-in-optimal}, the vendor's optimal investment under voluntary participation (VP) is the solution to the following optimization problem: 
% \begin{equation}\label{eq:static-optimization}
%     \begin{aligned}
%         \max_{x\geq 0}\;&\;G(x) \\ 
%         s.t.\;& \; G(x) \geq U^{\text{out},*}\quad && \text{(VP)}.
%     \end{aligned}
% \end{equation}
% Voluntary participation is satisfied if and only if the above optimization problem is feasible.
% }
%
To highlight the dependence of the vendor's opt-out utility on its risk attitudes, we will write $U^{\textrm{out},*}(\gamma)$ instead of $U^{\textrm{out},*}$ with liability loss taking the form given in Eqn \eqref{eq:liability-expression}. 
The next results follow immediately. 

%\com{TODO: formulate a optimization problem as 16.}\resp{I don't think we can formulate an interesting optimization problem for this section as 16 because this section studies when voluntary participation can be guaranteed (not the optimal investment under VP). This only involves the inequality $U^{\text{out},*}(\gamma)\leq U^{\text{in},*}$ and we are finding $\gamma$ that attains the equality.}

%\com{Ziyuan: when the vendor is indifferent between in and out, does it invest more (higher $x$) when out than in??} \resp{It is not immediately clear because it depends on which one ($x_L$ or $x_H$) is selected as the opt-in investment.}

% \begin{lemma}\label{lemma:gamma-mono}
%     $U^{\textrm{out},*}(\gamma)$ is monotonically decreasing in the risk attitude $\gamma$.
% \end{lemma}
%The following result follows directly from the lemma.

\begin{theorem}\label{thm:gamma-bar}
% (Criterion for Voluntary Participation)
% \begin{enumerate} 
    % \item[(1)] 
    There exists $\overline{\gamma}\in[0,\infty]$ such that for $\gamma\geq \overline{\gamma}$, $U^{\textrm{out},*}(\gamma)\leq U^{ *}$ and the vendor has the incentive to participate in the audit mechanism; for $\gamma< \overline{\gamma}$, $U^{\textrm{out},*}(\gamma)>U^{ *}$ and the vendor prefers to stay outside. Specifically, when $\overline{\gamma}=\infty$, the vendor never participates;  and when $\overline{\gamma} = 0$, the vendor always participates, regardless of the specification of the test function.
%    \com{Should these be the other way round? 0 is no risk aversion and therefore no participation and $\infty$ is extreme aversion and therefore always participate?} \resp{$\overline{\gamma}$ is the minimum risk aversion required for the vendor to participate. So 0 means any risk-averse vendor would participate, but $\infty$ means only infinitely risk-averse vendor would participate (this is clearly impossible, so in that case all risk-averse agent doesn't participate).}
    % \item[(2)] \rev{If $\overline{\gamma}<\infty$, then there exists $\overline{\gamma}'\in[\overline{\gamma},\infty]$ such that for $\overline{\gamma}\leq\gamma<\overline{\gamma}'$, the vendor invests more in the mechanism that opting out; for $\gamma>\overline{\gamma}'$, the vendor invests less in the mechanism than opting out. Specifically, when $\overline{\gamma}'=\overline{\gamma}$, the vendor always invests (weakly) less in the mechanism; and when $\overline{\gamma}'=\infty$, the vendor  always invests more in the mechanism.}
% \end{enumerate}
\end{theorem}
\rev{
\begin{proof}
    It is sufficient to show that $U^{\textrm{out},*}(\gamma)$ is monotonically decreasing in the risk attitude $\gamma$.
    Denote the opt-out utility function as $U^{\textrm{out}}(\gamma,x)$ to emphasize the role of risk aversion $\gamma$ in the specific liability loss form introduced by Eqn \eqref{eq:liability-expression}. First notice that $U^{\textrm{out}}(\gamma_1,x)\geq U^{\textrm{out}}(\gamma_2,x)$ for every $\gamma_1\leq \gamma_2$ and every $x$ as $C_L(\gamma, x)$ is increasing in $\gamma$ for every $x$. It directly follows that $U^{\textrm{out},*}(\gamma_1)\geq U^{\textrm{out},*}(\gamma_2)$ for every $\gamma_1\leq \gamma_2$. Thus, $U^{\textrm{out},*}(\gamma)$ is monotonically decreasing in $\gamma$. Define $\overline{\gamma}:=\inf\{\gamma\geq0\;|\; U^*>U^{\textrm{out},*}(\gamma)\}$ with the convention $\inf\;\emptyset:=\infty$. It can be verified that $\overline{\gamma}$ satisfies the desired properties.
\end{proof}
}

The value $\overline{\gamma}$ is the boundary risk attitude at which the vendor is indifferent between committing to the waiver/audit or not.  Above this level, the vendor is relatively risk-averse and, therefore, interested in participating and transferring its risk to the auditor. Below this level, the vendor is relatively risk-seeking and does not have an incentive to participate. 
% be interpreted as the boundary risk attitude that changes the vendor's behavior. When $\gamma>\overline{\gamma}$ or the vendor is very risk-averse, it is the vendor's interest to participate in the mechanism where the risk of liability loss is waived on its entirety (or transformed into the risk of auditing cost towards which the vendor holds a more neutral or even favoring attitude).  
% When $\gamma<\overline{\gamma}$ or the vendor is less risk-averse, it does not have the incentive to participate in the mechanism since the risk of liability loss is not too much a burden compared to the risk of the auditing cost. 

As $\overline{\gamma}$ is non-negative, participation is increased with a lower $\overline{\gamma}$. Below we show how the auditor can lower this value by adjusting its test threshold and noise. 
%\rev{If we assume a vendor's risk attitude $\gamma$ is drawn from a uniform distribution, we will use the minimum $\overline{\gamma}$ as a \textit{coverage metric} for a specific class of auditing regulations, where lower values indicate higher coverage capacities.} 

We will write the maximum opt-in utility as $U^{ *}(\delta,\sigma)$ to emphasize its dependence on the test parameters. For each pair of $(\delta,\sigma)$, we can calculate $\overline{\gamma}$ by solving  $U^{\textrm{out},*}(\overline{\gamma})=U^{ *}(\delta,\sigma)$ for $\overline{\gamma}$. We will similarly write it as $\overline{\gamma}(\delta,\sigma)$. Define the {\em coverage} of an audit mechanism with a fixed threshold $\delta$ as $\overline{\gamma}_\delta:=\inf_{\sigma\geq0}\overline{\gamma}(\delta,\sigma)$. Similarly, define the coverage associated with a fixed accuracy $\sigma$ as $\overline{\gamma}_\sigma:=\inf_{\delta\geq0}\overline{\gamma}(\delta,\sigma)$. The coverage of the audit mechanism where both $\delta$ and $\sigma$ are free variables is denoted as $\overline{\gamma}_0$.

\begin{theorem}\label{thm:gamma-delta-sigma}
    \begin{enumerate}
        \item[(1)] $\overline{\gamma}_0=0$, i.e., there exists an audit mechanism (a pair of test parameters) that ensures full coverage (for all vendor types).
        
        \item[(2)] For fixed $\sigma$ and $\delta_1\leq \delta_2$, $\overline{\gamma}(\delta_1,\sigma)\leq \overline{\gamma}(\delta_2, \sigma)$ and $\overline{\gamma}_\sigma=\overline{\gamma}(0,\sigma)$. Thus $\overline{\gamma}(\delta,\sigma)$increases in $\delta$ and the maximum coverage is reached when $\delta=0$. 
        %\com{Wouldn't this make the boundary an increasing function of $\delta$ rather than decreasing?? -- it decreases to 0 when the threshold decreases to 0, right?}\resp{That's right, sorry that was a mistake...} 
        However, this maximum is practically undesirable since a zero threshold means a non-investing vendor; a behavior that should not be encouraged.
        
        \item[(3)] $\overline{\gamma}_{\sigma_1}\leq\overline{\gamma}_{\sigma_2}$ for every $\sigma_1\leq\sigma_2$. This implies that higher accuracy increases the coverage by attracting less risk-averse vendors.
    \end{enumerate}
\end{theorem}
\rev{
\begin{proof}
    (1) Observe that $U^{\textrm{out},*}(\gamma)$ is bounded above by $R-C_X(0)-1=R-1$. Comparing to Eqn \eqref{eq:opt-utility}, it suffices to show that $U^*=W(0)>R-1$ for some test setup $(\delta,\sigma)$. This is equivalent to show that $$\sup_{\delta\geq 0,\sigma>0}\;\sup_{x\geq 0}\;G(x,\delta,\sigma) > R-1$$ where $G(x,\delta,\sigma)$ is the same function as Eqn \eqref{eq:G-def} with the function $p(\cdot)$ instantiated in Section \ref{sec:audit-quality}. We can switch the supremum, yielding
    \begin{equation*}
        \sup_{x\geq 0}\;\sup_{\delta\geq0,\sigma>0}\;G(x,\delta,\sigma) = \sup_{x\geq 0}\;R-C_X(x) = R
    \end{equation*}
    which completes the proof.

    (2) Since $U^{\textrm{out},*}(\overline{\gamma})=U^*(\delta,\sigma)$ and $U^{\textrm{out},*}(\gamma)$ is monotonically increasing in $\gamma$ by Theorem \ref{thm:gamma-bar}, it suffices to show that $U^*(\delta_1,\sigma)\geq U^*(\delta_2,\sigma)$ for any $\delta_1\leq\delta_2$ and fixed $\sigma$. Observe that the test function $p(x,\delta,\sigma):=1-\Phi((\delta-x)/\sigma)$ is decreasing in $\delta$ and thus we have $G(x,\delta_1,\sigma)\geq G(x,\delta_2,\sigma)$, where the result follows readily by taking $\sup$ over $x$ on both sides of the inequality.

    (3) We first show that $U^*(0,\sigma_1)\geq U^*(0,\sigma_2)$ for $\sigma_1\leq \sigma_2$. In the test form specified in the beginning of Section \ref{sec:audit-quality}, it is obvious that $p(x,0,\sigma_1)\geq p(x,0,\sigma_2)$ for every $x\geq 0$. Then, we have $G(x,0,\sigma_1)\geq G(x,0,\sigma_2)$ for every $x\geq 0$. Thus, by taking $sup$ over $x\geq 0$ on both sides, we obtain $U^*(0,\sigma_1)\geq U^*(0,\sigma_2)$.

    Using (2), we observe that $U^*(0,\sigma_1)\geq U^*(0,\sigma_2)\geq U^*(\delta,\sigma_2)$ for arbitrary $\delta\geq 0$. This implies the test function with parameters $(0,\sigma_1)$ yields higher opt-in utility than $(\delta,\sigma_2)$ with any specification of $\delta$. In light of the equality relationship $U^{\textrm{out},*}(\overline{\gamma})=U^*(\delta,\sigma)$ and the monotonicity of $U^{\textrm{out},*}(\cdot)$, we have $\overline{\gamma}(0,\sigma_1)\leq \overline{\gamma}(\delta,\sigma_2)$, $\forall\delta\geq0$, implying $\overline{\gamma}(0,\sigma_1)\leq \inf_{\delta\geq 0}\overline{\gamma}(\delta,\sigma_2)=\overline{\gamma}_{\sigma_2}$, which completes the proof.
\end{proof}
}

\begin{figure}[!htbp]
    \centering
    \includegraphics[width=0.8\linewidth]{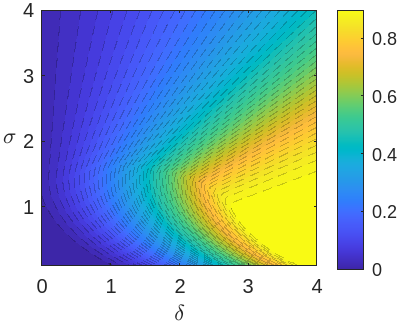}
    \caption{Contour plot of $\overline{\gamma}(\delta,\sigma)$ against various values of $\delta$ and $\sigma$. Other parameters are $c=1$, $\alpha=0.5$, $R=4$, $\mu_Z(x)=1/x$, and $\sigma_Z(x)=1.5/x$.}
    \label{fig:gamma-bar}
\end{figure}
Fig. \ref{fig:gamma-bar} shows some numerical simulations highlighting the above result. 
We make a similar observation that in terms of maximizing the mechanism's coverage or participation, it is once again best to have a highly accurate test but not a very strict/difficult one. 

\section{Designing a Dynamic Audit Scheme}
\label{sec:dynamic}

\rev{In this section, we will relax Assumption \ref{assump:threshold_audit} and consider a form of dynamic audit} and illustrate how to apply Theorem \ref{thm:opt-in-optimal} to characterize conditions under which higher security investments can be incentivized. \rev{We will do so by limiting ourselves to a special type of linear test function defined below.}

\subsection{A Linear Test Function}
Specifically, the test function considered in this section is a linear function truncated within $[0,1]$, given as follows: 
\begin{equation*}
    p(x)=\begin{cases}
        0 & 0\leq x< b \\
        x-b & b\leq x< b+1 \\
        1 &  x \geq b+1
    \end{cases}.
\end{equation*}
The parameter $b$ is the \textit{entrance value} interpreted as the minimum investment under which a pass is impossible. \rev{This number reflects the boundary of legal or societal tolerance of the software product quality and is akin to the idea of a {\em floor} advocated in \cite{lawfare_Dempsy24} (``the minimum legal standard of care for
software'')}.  
%which attempts to capture the notion that any investment below this level would imply an unacceptable product quality and, if granted entry to the market, would likely cause irreparable harm.} 
\rev{Similarly, the value $b+1$ represents a {\em ceiling}, or the so-called {\em safe harbor} also advocated in \cite{lawfare_Dempsy24} (``that
shields them from liability for hard-to-detect flaws''), the idea being that those who have made all conceivable effort should not be responsible for any residual risk and the associated losses.}
In between these two extreme values, the probability of passing the test is a linear function of the vendor's effort. Note that this formulation sets the slope of the linear test function to $1$, which is without loss of generality since other linear forms can always be transformed to the above formulation by re-scaling the $x$-axis. 

\rev{The above linear test function is a special case, and therefore, all previous results hold when it is used independently in successive tests. However, by focusing on a specific form of test, we are able to study a more complex, dynamic audit mechanism, whereby successive tests may employ a different parameter $b$, as we detail below.}

For the rest of this section, we will occasionally write $p(x)$ as $p_b(x)$ to emphasize its dependence on $b$. Observe that for $b'>b$, we have $p_{b'}<p_b$, meaning that $p_{b'}$ is harder than $p_b$, i.e., adopting a larger value of $b$ leads to a harder test to pass. 

The $G(\cdot)$ function in Lemma \ref{lemma:explicit-w} can be \rev{re-written} as follows
\begin{equation}\label{eq:Gx-linear}
    G(x) = \begin{cases}
        -cx & 0\leq x < b \\
        -cx + R\cdot\frac{x-b}{1-\alpha + \alpha x - \alpha b} & b\leq x < b+1 \\
        R-cx & x\geq b+1
    \end{cases}.
\end{equation}
We will write $G^b(x)$ instead of $G(x)$ in the rest of this section to emphasize its dependence on the entrance value $b$.

\subsection{Static Audit with the Linear Test Function}

For comparison, we will first examine the extent to which a static audit with test function $p_b$ \rev{(i.e., $p_b$ is applied independently following each failed test)} can incentivize security investments. From the auditor's point of view, this can be formulated as an optimization problem that maximizes the vendor's security investment over the parameter $b$:
% \begin{equation}\label{eq:static-optimization}
% \begin{aligned}
%     % \max_{b\geq 0}\;&\;x \\
%     % s.t.\;&\; G^b(x) \geq G^b(y),\;\;\forall y\geq 0 \\
%     % &\; x\geq 0
% \end{aligned}
% \end{equation}
\rev{
\begin{equation}\label{eq:static-optimization}
    \begin{aligned}
        \max_{b\geq 0}\;&\;x \\ 
        s.t.\;&\; x\in\arg\max_{y\geq 0}\;G^b(y)\quad && \text{(IC)} \\
        & \; G^b(x) \geq 0\quad && \text{(VP)}
    \end{aligned}
\end{equation}
where the first and the second constraints describe incentive compatibility (IC) and voluntary participation (VP), respectively.} 
\rev{Notice that we intentionally set the optimal outside utility as zero in the VP constraint without loss of generality -- since the vendor's outside utility is independent of the audit, our results can be easily adapted to the more general case by offsetting $G^b(\cdot)$ by the optimal outside utility.} 
Denote the maximizer to \eqref{eq:static-optimization}, \rev{if it exists,} as $b^*$ and the corresponding vendor's investment as $x(b^*)$. We will refer to $x(b^*)$ as the {\em incentivizable investment}.
\begin{proposition}\label{prop:static-solution}
    The solution to Problem \eqref{eq:static-optimization} is given as follows:
    \begin{enumerate}[(i)]
        \item If $0\leq \frac{R}{c}<1-\alpha$, $b^*$ is arbitrary and $x(b^*)\equiv 0$. \label{prop:static-solution-1}
        % This means the vendor always receives zero opt-in utility regardless of the test function $p_b$.
        \item If $1-\alpha \leq \frac{R}{c} < \frac{1}{1-\alpha}$,  $b^*=\frac{\left(\sqrt{R}-\sqrt{(1-\alpha)c}\right)^2}{\alpha c}$ and $x(b^*)=\frac{R - \sqrt{(1-\alpha)Rc}}{\alpha c}$.\label{prop:static-solution-2}
        \item If $\frac{R}{c}\geq \frac{1}{1-\alpha}$, $b^*=\frac{R}{c}-1$ and $x(b^*) = \frac{R}{c}$.\label{prop:static-solution-3}
    \end{enumerate}
\end{proposition}
\rev{
\begin{proof} 
Observe: 
\begin{enumerate}[(a)]
    \item $G^b(x)$ decreases respectively over the intervals $[0,b)$ and $[b+1,\infty)$.
    \item Write $f(x)=-cx+R\frac{x-b}{1-\alpha+\alpha (x-b)}$ (the expression of $G(x)$ when $x\in[b,b+1)$). Then, $f(x)$ is increasing when $x\leq \overline{x}$ and decreasing when $x\geq \overline{x}$, where $\overline{x}:=b-\frac{1-\alpha}{\alpha} + \sqrt{\frac{(1-\alpha)R}{\alpha^2c}}$ and $f(\overline{x})=\alpha^{-1}\left[\left(\sqrt{R}-\sqrt{c(1-\alpha)}\right)^2-\alpha bc\right]$. 
\end{enumerate} 
For case \eqref{prop:static-solution-1}, $G^b(\cdot)$ is strictly decreasing over $\mathbb{R}_+$ given observation $(a)$ and that $\overline{x}<b$ implies $G^b(\cdot)$ decreases over $[b,b+1)$. Thus, the IC constraint in \eqref{eq:static-optimization} implies $x=0$. For case \eqref{prop:static-solution-2}, we have $\overline{x}\in[b,b+1)$. Thus, $\overline{x}$ is a local maximum of $G^b(\cdot)$ (the other local maximum is attained at $x=0$). Thus, \eqref{eq:static-optimization} reduces to
\begin{equation*}
        \max_{b\geq 0}\;\;\overline{x}\quad s.t.\;G^b(\overline{x})\geq 0,
\end{equation*}
where the solution is obtained by setting $G^b(\overline{x})=0$ as $\overline{x}$ is increasing in $b$ and $G^b(\overline{x})=f(\overline{x})$ is decreasing in $b$.
For case \eqref{prop:static-solution-3}, we have $\overline{x}\geq b+ 1$. Thus, $G^b(\cdot)$ is increasing over $[b,b+1)$ and $b+1$ is a local maximum. Similarly, the solution is obtained by solving
\begin{equation*}
    \max_{b\geq0}\;\;b+1\quad s.t.\;G^b(b+1) \geq 0,
\end{equation*}
which is a simple linear program.
\end{proof}
}

The partition criterion $R/c$ can be viewed as the return on security investment (RoSI) \cite{sonnenreich2005} or profitability of the vendor under the audit mechanism. In case \eqref{prop:static-solution-1} where the vendor has low ROSI, there is no way to incentivize the vendor by manipulating the entrance value. This means vendors falling into this case would not voluntarily participate in the audit mechanism in the first place. In case \eqref{prop:static-solution-2}, all incentivizable investments are (tightly) upper bounded by $x(b^*)$. \rev{Let $R/c$ be the {\em (security) investment capacity}, i.e., the vendor spends all revenue on security.} The following inequality implies that the incentivizable investment $x(b^*)$ is at most $\frac{1}{4\alpha}$ less than the investment capacity:
\begin{equation*}
    \begin{aligned}
        \frac{R}{c}-\overline{x} &= \frac{1}{\alpha}\left(\sqrt{\frac{(1-\alpha)R}{c}}-\frac{(1-\alpha)R}{c}\right) \\ &= \frac{1}{\alpha}\left[\frac{1}{4}-\left(\sqrt{\frac{(1-\alpha)R}{c}}-\frac{1}{2}\right)^2\right] \leq B(\alpha),
    \end{aligned}
\end{equation*}
where $B(\alpha) = \frac{1}{4\alpha}$ when $\alpha \geq \frac{1}{2}$ and $B(\alpha) = 1-\alpha$ when $\alpha < \frac{1}{2}$. Notice that the bound of this difference $B(\alpha)$ is decreasing in $\alpha$. The intuition is that when the vendor is short-sighted/myopic ($\alpha\searrow0$), it is very sensitive to the hardness of the test (or audit as it is static), as harder ones are usually associated with later payoffs, a situation devalued by shortsighted vendors. Thus, $b^*$ should be set conservatively for such vendors, and higher investments become less likely. At the other end, when the vendor is far-sighted ($\alpha\nearrow1$), it is more willing to invest close to $R/c$ since later revenue is not affected too much by devaluations.
In case \eqref{prop:static-solution-3} where the vendor has high ROSI, the investment capacity is attainable/incentivizable by setting $b=R/c-1$. 
%\com{Another observation is that when the vendor is shortsighted, case (ii) is a very small range, whereas if it is farsighted, case (ii) is a much bigger range. I wonder if there are implications associated with this...} 
\rev{Furthermore, %Proposition \ref{prop:static-solution} indicates that only case (ii) allows for the potential improvement of incentivizable investments. 
when the vendor is short-sighted, the range in case (ii) is very narrow, suggesting limited chances for further incentivization. Conversely, if the vendor is far-sighted, the range in case (ii) is significantly broader, indicating greater potential for further incentivization.}

\subsection{A Dynamic (Two-Step) Audit with the Linear Test Function}\label{sec:two-step-dynamic}

We next explore whether the incentivizable investment can be further improved by prepending an extra (but different) test $p_{b'}$ before the static audit with test $p_b$. 
% \rev{We denote the entrance value of the additional audit as $b_e$ (resp. $b_h$) if it is easier, i.e., $b_e<b$ (resp. harder, i.e, $b_h>b$), than the following repeated audits and the associated test function as $p_{b_e}$ (resp. $p_{b_h}$).} 
We introduce the following notations to emphasize their dependence on the audit: $U^{b,*}(x):=\max_{y\geq x}\;G^b(y)$ is the vendor's maximum utility provided its investment is not less than $x$; $U^{b',b}(x):=-(1-\alpha + \alpha p_{b'}(x))C_X(x) + p_{b'}(x)R + \alpha(1-p_{b'}(x))U^{b,*}(x)$ is the vendor's utility in the proposed audit provided that it acts optimally after and including the second stage. The equivalent optimization formulation for dynamic audits is 
% \begin{equation}\label{eq:dynamic-optimization}
% \begin{aligned}
%     \max_{b,\;b'\geq 0}\;&\;x \\
%     s.t.\;&\; U^{b',b}(x) \geq U^{b',b}(y),\;\;\forall y\geq 0 \\
%     &\; x\geq 0.
% \end{aligned}
% \end{equation}
\rev{
\begin{equation}\label{eq:dynamic-optimization}
\begin{aligned}
    \max_{b,\;b'\geq 0}\;&\;x \\
    s.t.\;&\; x\in\arg\max_{y\geq 0}\;U^{b',b}(y)\quad && \text{(IC)} \\
    &\; U^{b',b}(x)\geq 0\quad &&\text{(VP)}.
\end{aligned}
\end{equation}
}
Denote the maximizer to \eqref{eq:dynamic-optimization} as $\overline{b}$ and $\overline{b}'$, with the incentivizable investment in this setup, i.e., the optimal objective to \eqref{eq:dynamic-optimization}, as $x(\overline{b},\overline{b}')$.

\begin{lemma}\label{lemma:dynamic-impossibility}
    The incentivizable investment cannot be improved by dynamic audits if $\frac{R}{c}< 1-\alpha$ or  $\frac{R}{c} \geq \frac{1}{1-\alpha}$.
\end{lemma}
\begin{proof}[Proof Sketch]
    The proof involves evaluating $U^{b,*}(\cdot)$ and $U^{b,b'}(\cdot)$ for various conditions of ROSI. Then, analyze the monotonicity of $U^{b,b'}(\cdot)$ over every partition of $\mathbb{R}_+$ based on the relationship between $b$, $b'$, $b+1$, and $b'+1$. The detailed proof is shown in \ref{app:dynamic-impossible}.
\end{proof}
Lemma \ref{lemma:dynamic-impossibility} states that only case \eqref{prop:static-solution-2} in Proposition \ref{prop:static-solution} can possibly be improved by the dynamic audit setup. Intuitively, vendors in case \eqref{prop:static-solution-1} are too cost-inefficient to be incentivized by any mechanism, and those in case \eqref{prop:static-solution-3} have already invested as much as they could possibly afford. 
\rev{Notice that Lemma \ref{lemma:dynamic-impossibility} does not assume the relationship between $b$ and $b'$. In the rest of this subsection, we separately examine the cases of $b'<b$ (an easy test followed by a sequence of harder ones) and $b'>b$ (a hard test followed by a sequence of easier ones) in Proposition \ref{prop:dynamic-solution} and \ref{prop:dynamic-impossible-harder}, respectively.}

% To completely solve \eqref{eq:dynamic-optimization}, we need to evaluate $U^{b',b}$ over 5 pieces of the positive real line divided by the four numbers $b$, $b+1$, $b'$, and $b'+1$. Additionally, there are four different types of divisions based on the relationship between $b$ and $b'$, resulting in a total of $4\times 5=20$ evaluations. For simplicity, we will only present and analyze two representative solutions in the following propositions.
\begin{proposition}\label{prop:dynamic-solution}
    Suppose $1< \frac{R}{c}< \frac{1}{1-\alpha}$. The test functions characterized by the following entrance values constitute a solution to \eqref{eq:dynamic-optimization}:
    \begin{equation*}
        \overline{b} = \frac{R}{c} + \left(\sqrt{\frac{R}{\alpha c}}-\sqrt{\frac{1-\alpha}{\alpha}}\right)^2\quad\text{and}\quad \overline{b}' = \frac{R}{c} - 1
    \end{equation*}
    with the incentivizable investment $x(\overline{b},\overline{b}')=\frac{R}{c}$.
\end{proposition}
\rev{
\begin{proof}[Proof Sketch]
    We first show $U^{b,b'}(x)=-cx+p_{b'}(x)R$ when $x\leq L:=b-\Big(\sqrt{\frac{R}{\alpha c}}-\sqrt{\frac{1-\alpha}{\alpha}}\Big)^2$. Supposing $b'<b'+1\leq L$ and $\frac{R}{c}>1$, we show $U^{b,b'}(\cdot)$ is increasing over $[b',b'+1)$ and decreasing over $[0,b')$ and $[b'+1,\infty)$ respectively. The idea is to choose $b$ and $b'$ such that the vendor's maximum utility is attained at $b'+1$. Problem \eqref{eq:dynamic-optimization} restricted to this case is equivalent to
    \begin{equation*}
        \begin{aligned}
            \max_{b',\;b\geq 0}\;&\; b'+1 \\
            s.t.\;&\;U^{b',b}(b'+1) = R - c(1+b')\geq 0 \\
            &\;b'+1\leq L.
        \end{aligned}
    \end{equation*}
    The first constraint satisfies both IC and VP since the other local maximum, in this case, is $x=0$ with $U^{b',b}(0)=0$. This is a linear program and can be easily solved. The proof is completed by noticing that $x(\overline{b},\overline{b}')=\frac{R}{c}$ is indeed the global maximum as it equates to the investment capacity. Detailed proof is provided in Appendix \ref{app:dynamic-incentive}. 
\end{proof}
}
\rev{The dynamic audit suggested by Proposition \ref{prop:dynamic-solution} consists of an easier test in the first step followed by an infinite sequence of hard ones. The idea behind the mechanism is to encourage the vendor to always try to pass the audit in the first step.} Thus, we are able to improve the incentivizable investment by explicitly setting the first test that induces the investment capacity $R/c$.

However, a harder test followed by an infinite sequence of easier ones is not able to incentivize higher investments, as described by the following proposition.
\begin{proposition}\label{prop:dynamic-impossible-harder}
    Suppose we add a constraint $b+\varepsilon \leq b' \leq b+1$ to the optimization problem \eqref{eq:dynamic-optimization} with $\varepsilon>0$ emphasizing the minimum difference between the two tests. When $1-\alpha < \frac{R}{c} < \frac{1}{1-\alpha^2}$, the solution $(\overline{b},\overline{b}')$ to the modified optimization problem exists, 
    \begin{equation*}
        \overline{b} = \frac{1-\alpha}{\alpha}+\frac{R}{\alpha c} - 2\sqrt{\frac{(1-\alpha)R}{\alpha^2 c} + \frac{(1-\alpha)R}{\alpha c}\varepsilon},\;\;\overline{b}'=\overline{b} + \varepsilon,
    \end{equation*}
    and the incentivizable investment is
    \begin{equation*}
        x(\overline{b},\overline{b}') = \overline{b}-\frac{1-\alpha}{\alpha} + \sqrt{\frac{(1-\alpha)R}{\alpha^2 c} + \frac{(1-\alpha)R}{\alpha c}\varepsilon}.
    \end{equation*}
When $\frac{1}{1-\alpha^2}\leq R < \frac{1}{1-\alpha}$, the above audit is still optimal given that $\varepsilon$ is sufficiently small, i.e., $\varepsilon<\frac{c}{\alpha(1-\alpha)R}-\frac{1}{\alpha}$. Otherwise, the solution to the modified optimization problem, denoted alternatively as $(\Tilde{b},\Tilde{b}')$, is given by
\begin{equation*}
    \Tilde{b} = \frac{R - (1+\alpha)c}{\alpha c}\quad\text{and}\quad \Tilde{b}' = \frac{R}{\alpha c} + \frac{c}{\alpha (1-\alpha)R} - \frac{2+\alpha}{\alpha}
\end{equation*}
with the incentivizable investment $x(\Tilde{b},\Tilde{b}')=\frac{R-c}{\alpha c}$.
Importantly, both $x(\overline{b},\overline{b}')$ and $x(\Tilde{b},\Tilde{b}')$ are less than $\frac{R-\sqrt{(1-\alpha)Rc}}{\alpha c}$, the incentivizable investment for static audits.
\end{proposition}
\rev{
\begin{proof}[Proof Sketch]
    We first show that when $b<b'\leq b+1$, the vendor's optimal investment can only take three values: $0$, $\overline{x}_h$, or $b+1$. Finding the incentivizable investment is sensible only when the vendor's investment takes the latter two values. Then, we separately discuss the following subsets of the constraint set: $\overline{x}_h<b'$, $b'\leq \overline{x}_h\leq b+1$, and $\overline{x}_h\geq b+1$. The key observation is that these cases correspond respectively to the three optimal investments above. We find the incentivizable investment for each case and compare them for the highest. Detailed proof is provided in Appendix \ref{app:dynamic-incentive}.
\end{proof}
}
\rev{Proposition \ref{prop:dynamic-impossible-harder} implies that it is impossible to increase the incentivizable investment when using a harder test followed by an infinite sequence of easier tests; this type of dynamic audit cannot do better than the static audit.
The reason is two-fold. 
First, the vendor is more likely to opt out if the initial test becomes harder, so $b$ needs to be sufficiently small in order to not cause the vendor to opt out; this limits our ability to incentivize a higher investment. 
%
%%%%%BG ready until here!

Secondly, it is difficult for the vendor to see the value of the easier tests in the future due to its shortsightedness; this is regardless of how short-sighted it is, as long as it is ($\alpha < 1$). 
%So, future easier audits are not enough to incentivize a higher investment, making this type of dynamic audit unable to do better than the static audit.
%Proposition \ref{prop:dynamic-impossible-harder} shows that this inability exists regardless of the level of the vendors' shortsightedness, as long as they are ($\alpha < 1$).
In fact, when the vendor is not shortsighted at all ($\alpha =1$), and we set $b=\frac{R}{c}$, any participating vendor would invest up to $\frac{R}{c}$ since it will receive full revenue with probability one.
% This can be justified by decreasing $\alpha$ (alleviating shortsightedness) in Eq. \eqref{eq:dynamic-harder-bound}, where the bound converges to the theoretical maximum $\frac{R}{c}$ when $\alpha$ converges to $1$.
}
%\com{1: Prop 5.4 does not require $\alpha$ to be small, correct, just the value of the bounds depends on it? So the intuition is that no matter how shortsighted the vendor is or is not, future easier audits are not enough to get them to invest more by adding a harder audit at the beginning? 2: what does it mean that the initial harder audit scares off the vendor: that they are more likely to not voluntarily participate?}
%\resp{1. Yes, this is correct. A harder audit before easier ones cannot do better than the static audit regardless of the value of $\alpha$. But $\alpha$ has to be less than 1 for this result to hold. 2. Yes, this was what I meant initially. The vendors are more likely to opt out if the initial audit becomes harder, so to maintain them inside the mechanism, we should choose $b$ conservatively, which limits our ability to incentivize a higher investment. However, I think this phrase "scare off" could be confusing so I reframed the explanation above a little.}

%\com{I think it might be clearer to use $b$ to denote the infinite sequence of audits from step 2 onwards and use something like $b_1$/$b_2$ or $b_e$/$b_h$ to denote the easier or harder audit we add in step 1?}\resp{Thanks! I think it is a good idea, but the the formulation and Lemma 5.2 applies to all test functions. How about using $b'$ before Proposition 5.3 and introduce $b_h/b_e$ in Proposition 5.3 and 5.4?}

\subsection{The General Case: Multi-Step Audit with Arbitrary Tests}\label{sec:general-dynamic}

%\com{We could also move a couple of the more general results on dynamic audits here from the appendix?}

Consider a general non-static audit $A=\{p_n(\cdot)\}_{n=0}^\infty$ where $p_n(\cdot)$ satisfies Assumption \ref{assump:threshold_audit}. Denote the vendor's optimal (ex-ante) utility under $A$ as $U^{A,*}$. Let $U_n^{A}(x)$ be the vendor's utility function under audit $A$ starting from the $n$-th step (the step with test function $p_n(\cdot)$). Define $U_n^{A,*}(x):=\max_{y\geq x}\;U_n^{A}(y)$. Then, the optimal utility can be expressed as $U^{A,*}=U_0^{A,*}(0)$. Solving $U^{A,*}$ directly can be intractable. We thus examine an approximation with a sequence of ``truncated'' versions of $A$. Define $A_k:=(p_0,p_1,\dots,p_{k-1},p_k,p_k,\dots)$ as the audit whose test functions are the same as $A$ for the first $k+1$ steps ($0$-th to $k$), but then repeats $p_k$ indefinitely after the $k$-th step. Our goal is to show that $U^{A}(\cdot)$ can be sufficiently well-approximated by the sequence $\{U_0^{A_k}(\cdot)\}_{k=1}^\infty$, and that the vendor's optimal investment under $A_k$ converges to its optimal investment under $A$. As a special case, we also examine how well a two-step audit approximates an arbitrary one. In the following results, we assume all maximums are attainable.

As the first step, we examine how the change of one test in a sequence under the audit $A$ affects the vendor's utility. Replace the $m$-th function in $A$, i.e., $p_m$, by a different function $q_m$ and denote the resulting audit as $B$. The following lemma bounds the difference between the two.

\begin{lemma}\label{lemma:one-change}
    When $q_m>p_m$, $U^{B,*}\geq U^{A,*}$; when $q_m<p_m$, $U^{A,*}\leq U^{B,*}$. 
    Moreover, $|U^{B,*} - U^{A,*}|\leq \alpha ^m\Delta R$, where $\Delta:=||q_m-p_m||_\infty\leq 1<\infty$.
\end{lemma}
\begin{proof}
    We only show the case for $q_m>p_m$. The other direction readily follows from similar arguments.

    According to the dynamic programming principle, we obtain a recursive relationship between the functions $U^A_n(\cdot)$ and $U^A_{n+1}(\cdot)$ for each $n$:    \begin{equation*}
        \begin{aligned}
            U_n^A(x) = - (1-\alpha + \alpha p_n(x)) C_X(x) + p_n(x)R \\+ \alpha (1-p_n(x)) U_{n+1}^{A,*}(x)~.    
        \end{aligned}
    \end{equation*}
    It is straightforward to show that $U_{m+1}^{A,*}(\cdot)=U_{m+1}^{B,*}(\cdot)$ because all functions after (not including) the $(m+1)$-th step are the same for both $A$ and $B$.
    At the $(m+1)$-th step, we observe
    \begin{equation*}
        \begin{aligned}
            &U_m^{B}(x) - U_m^{A}(x) = \\ &\;\;\;\;(q_m(x)-p_m(x))\big[R - \alpha \big(C_X(x) + U_{m+1}^{A,*}(x)\big)\big]\geq 0.
        \end{aligned}
    \end{equation*}
    The inequality follows from Lemma \ref{lemma:bdd}. Thus, we have $U_m^B(x)\geq U_m^A(x)$ for every $x$ and thus $U_m^{B,*}(x) \geq U_m^{A,*}(x)$. 
    
    At the $m$-th step, we have 
    \begin{equation*}
        \begin{aligned}
            &U_{m-1}^{B}(x) - U_{m-1}^{A}(x) = \\ &\;\;\;\;\alpha (1-p_{m-1}(x))\big[U_m^{B,*}(x)-U_m^{A,*}(x)\big] \geq 0,
        \end{aligned}
    \end{equation*}
    which again implies $U_{m-1}^{B,*}(x) \geq U_{m-1}^{A,*}(x)$. Apply the same arguments repeatedly until we arrive at $U_{0}^{B}(x) - U_{0}^{A}(x)\geq 0\implies U_{0}^{B,*}(x) \geq U_{0}^{A,*}(x) \iff U^{B,*} \geq U^{A,*}$.

    For the upper bound of $|U^{B,*} - U^{A,*}|$, we apply Lemma \ref{lemma:bdd} to the above two inequalities, yielding
    \begin{equation*}
        |U_m^{B,*}(x) - U_m^{A,*}(x)|\leq\max_{z\geq x}\;|U_m^{B}(z) - U_m^{A}(z)|\leq \Delta R
    \end{equation*}
    and 
    \begin{equation*}
        \begin{aligned}
            |U_{m-1}^{B,*}(x) - U_{m-1}^{A,*}(x)|&\leq \max_{z\geq x}\;|U_{m-1}^{B}(z) - U_{m-1}^{A}(z)| \\&\leq \alpha \Delta R.
        \end{aligned}
    \end{equation*}
    Apply repeatedly and obtain $|U^{B,*} - U^{A,*}|=|U_0^{B,*}(0) - U_0^{A,*}(0)|\leq \alpha^m \Delta R$ where we used that $U_0^{A,*}(x)$ and $U_0^{B,*}(x)$ are decreasing, the proof of which is exactly the same as Lemma \ref{lemma:mono-w}.
\end{proof}

The exponential upper bound in Lemma \ref{lemma:one-change} indicates how the change of future tests is reflected in the vendor's initial utility. The vendor's experience of this change decays exponentially in the vendor's shortsightedness $\alpha$. This also explains why a harder initial test followed by a sequence of easier ones cannot incentivize a higher investment than the static counterpart. Lemma \ref{lemma:one-change} also leads to a monotonicity property for two audits $A$ and $B$: if every function in $A$ is larger (resp. smaller) than that in $B$, then the vendor's optimal utility is higher (resp. lower) in $A$ than $B$.

The proof of Lemma \ref{lemma:one-change} also implies a $\sup$-norm bound on the vendor's utility functions: $||U^B_k-U^{A}_k||_\infty\leq \alpha ^{m-k}\Delta R$ for every $k\leq m$. This indicates that the utility functions under the two audits get close to each other uniformly as $m\to\infty$. This enables us to approach the optimal vendor's utility in any audit, as well as its optimal initial investment, by constructing sequences that converge to the desired audit.

\begin{theorem}\label{thm:approximation}
    The sequence of vendor's utility functions under audit $A_k$, i.e., $\{U^{A_k}_0\}_{k=0}^\infty$, converges uniformly to $U^{A}_0$ as $k\to\infty$ with sup-norm error $\frac{\alpha^{k+1}R}{1-\alpha}$. As a special case, when $k=2$, 
    \begin{equation*}
        ||U_0^A-U_0^{A_2}||_\infty\; \leq \; \frac{\alpha^2R}{1-\alpha}.
    \end{equation*}
    
    Moreover, if the maximizers of $U_0^A$ and $U_0^{A_k}$'s are unique, denoted respectively as $x^*$ and $x^*_k$, then $x^*_k$ converges to $x^*$ as $k\to\infty$.
\end{theorem}

\begin{proof}
    Notice $A$ can be constructed from $A_k$ by modifying test functions after $k$-th location one at a time. This implies a telescoping sum of errors, i.e., 
    \begin{equation*}
        ||U_0^{A_k}-U_0^A||_\infty \leq \sum_{j=k+1}^\infty \alpha ^j R = \frac{R\alpha ^{k+1}}{1-\alpha},
    \end{equation*}
    where we utilized the fact that $\Delta\leq 1$.

    Now let $\{x_{k_i}^*\}_{i=0}^\infty$ be any convergent subsequence of $\{x^*_k\}_{k=0}^\infty$. Suppose the limit of $\{x_{k_i}^*\}_{i=0}^\infty$, denoted as $x'$, is different from $x^*$. By uniqueness, $U^A_0(x^*)-U_0^A(x') =: \gamma > 0$. Then, by continuity and uniform convergence of $U_0^{A_k}$, there exists $M>0$, s.t., $\forall i > M$, we have $|U_0^{A_{k_i}}(x_{k_i}^*)-U^{A}_0(x')|<\gamma/2$, which implies $|U_0^{A_{k_i}}(x_{k_i}^*)-U^{A}_0(x^*)|>\gamma/2$ contradicting to the uniform convergence. Thus, we have shown every convergent subsequence converges to the same $x^*$, which implies $\{x^*_k\}_{k=0}^\infty$ converges to $x^*$.
\end{proof}

\rev{
Theorem \ref{thm:approximation} implies that there always exists a two-step audit that uniformly (in sup-norm) approximates any general audit up to an error of $\frac{\alpha^2R}{1-\alpha}$.
This bound is small when $\alpha$ is small, i.e., the vendor is short-sighted, and trivially large (approaches $\infty$) when $\alpha$ is close to $1$, i.e., the vendor is far-sighted. This result supports the intuition that few-step (shallow) test functions are generally sufficient for regulating shorted-sighted vendors, which potentially constitute a significant fraction in practical settings. This also provides additional justification for analyzing the two-step audits in the previous subsection.
A closer approximation can be obtained by replacing the two-step audit, e.g., $A_2$, with a $k$-step audit, e.g., $A_k$. Though harder than two-step audits, the problem with $A_k$ is indeed solvable with a $k$-step backward induction. Notice that $k$ does not need to be prohibitively large due to the exponential decrease in the error, suggesting the practical value of such an approximation.
}

\section{Conclusion}\label{sec:discuss}

In this work, we studied a (software security) audit problem motivated by the proposed liability waiver mechanism and examined its innate ability to incentivize desirable secure software development practices. This is formulated as an MDP problem with a full characterization of the properties of an optimal effort strategy by the (software) vendor. The most interesting result is that there can be many equally optimal strategies, in the form of a sequence of investments over time, that reflect very different attitudes and trade-offs on the part of the vendor, from low investment and slow return to high investment and rapid return. 
We also examined how the test parameters affect the vendor's participation incentive and showed that, in general, 
%We showed that the threshold audit mechanism could incentivize participation while inducing high security investment at the same time. 
an accurate (low noise) but not very strict (difficult) test is the most effective.   
\rev{Finally, we demonstrated how a two-step audit can increase incentivizable investments and how a general dynamic audit can be reasonably approximated by a two-step or finite-step audit.}

\rev{
There are a number of interesting future directions to pursue. This includes the comprehensive comparison with liability insurance mentioned earlier and potentially a constrained optimization framework that maximizes the worst-case audit coverage capacity subject to a certain minimum effort requirement.
} 

% \section{Conclusion}
% \label{sec:conclusion}
% In this work, we studied a (software security) audit problem motivated by the proposed liability waiver mechanism and examined its innate ability to incentivize desirable secure development practices. This is formulated as an MDP problem with a full characterization of the properties of an optimal effort strategy by the (software) vendor. The most interesting result is that there can be many equally optimal strategies, in the form of a sequence of investments over time, that reflect very different attitudes and tradeoffs on the part of the vendor, from low investment and slow return to high investment and rapid return. 
% We also examined how the audit parameters affect the vendor's participation incentive and showed that, in general, 
% %We showed that the threshold audit mechanism could incentivize participation while inducing high security investment at the same time. 
% an accurate (low noise) but not very strict (difficult) audit is the most effective.   

\bibliographystyle{IEEEtran}
\bibliography{otherbibs}

\appendix[Complete Proofs in Subsection \ref{sec:two-step-dynamic}]
The proof of Lemma \ref{lemma:dynamic-impossibility}, Proposition \ref{prop:dynamic-solution}, and Proposition \ref{prop:static-solution-3} all rely heavily on the explicitly expression of $G^b(x)$, $U^{b,*}(x)$, and $U^{b',b}(x)$. We first ignore the domain constraint and study $U^{b,*}(x)$ and $U^{b',b}(x)$ over the entire real line. Then, we apply the restriction $x\geq0$ to the obtained results. For ease of notation, we will use the abbreviation $\overline{\alpha}:=\frac{1-\alpha}{\alpha}$.

We introduce two important functions that will appear in the expression of $U^{b',b}(x)$ later on. Define
\begin{equation*}
    \begin{aligned}
        &g(x)=-cx+R+\\&\;\;\;\;(1+b'-x)(\alpha c x + (1-\alpha) c - 2\sqrt{(1-\alpha)Rc} - \alpha bc),
    \end{aligned}
\end{equation*}
which is a quadratic function with negative second-order coefficient. Its first order derivative equals
\begin{equation*}
    \begin{aligned}
    g'(x) = -2\alpha c x + \alpha c(b' + b) + 2\sqrt{(1-\alpha)Rc} - 2(1-\alpha)c.
    \end{aligned}
\end{equation*}
So, the axis of symmetry of $g(x)$, denoted as $\overline{x}_g$, is
\begin{equation*}
    \overline{x}_g = \frac{b'+b}{2} - \overline{\alpha} + \sqrt{\frac{\overline{\alpha}R}{\alpha c}}.
\end{equation*}
$g(x)$ is increasing when $x\leq \overline{x}_g$ and is decreasing when $x\geq \overline{x}_g$.
Define 
\begin{equation*}
    h(x) = -cx + \frac{x - (1-\alpha )b' - \alpha b}{1-\alpha + \alpha (x-b)}R
\end{equation*}
Applying first-order condition to $h(x)$, we see that $h(x)$ attains its maximum at
\begin{equation*}
    \overline{x}_h=b-\overline{\alpha}+\sqrt{\frac{\overline{\alpha}R}{\alpha c} + \frac{(1-\alpha)R}{\alpha c}(b'-b))}.
\end{equation*}
$h(x)$ is increasing when $x\leq \overline{x}_h$ and decreasing  when $x\geq \overline{x}_h$. We also derive that
\begin{equation*}
    \begin{aligned}
    h(\overline{x}_h) &= -c\cdot\left(b - \overline{\alpha} - \frac{R}{\alpha c} + 2\sqrt{\frac{\overline{\alpha}R}{\alpha c} + \frac{(1-\alpha)R}{\alpha c}(b'-b)}\right) \\ &= -c\cdot\left(\overline{x}_h - \frac{R}{\alpha c} + \sqrt{\frac{\overline{\alpha}R}{\alpha c} + \frac{(1-\alpha)R}{\alpha c}(b'-b)}\right).
    \end{aligned}
\end{equation*}
Notice that $\overline{x}_h$ (so is $h(\overline{x}_h)$) is generally not linear or convex.

\subsection{Proof of Lemma \ref{lemma:dynamic-impossibility}}\label{app:dynamic-impossible}
\begin{lemma}[Restatement of Lemma \ref{lemma:dynamic-impossibility}]
    The incentivizable investment cannot be improved by dynamic audits if $\frac{R}{c}< 1-\alpha$ or  $\frac{R}{c} \geq \frac{1}{1-\alpha}$.
\end{lemma}
\begin{proof}
    Recall the expression of $G^b(\cdot)$:
    \begin{equation*}
    G^b(x) = \begin{cases}
        -cx & 0\leq x < b \\
        -cx + R\cdot\frac{x-b}{1-\alpha + \alpha x - \alpha b} & b\leq x < b+1 \\
        R-cx & x\geq b+1
    \end{cases}.
\end{equation*}
    When $\frac{R}{c}<1-\alpha$, it is easy to show that $G^b(x)$ is strictly decreasing. Thus, by definition, $U^{b,*}(x)=\max_{z\geq x}\;G^b(z)=G^b(x)$. Plug it into the expression $U^{b',b}(x):=-(1-\alpha + \alpha p_{b'}(x))C_X(x) + p_{b'}(x)R + \alpha(1-p_{b'}(x))U^{b,*}(x)$ and obtain:
    \begin{equation*}
        U^{b',b}(x) = \begin{cases}
            -cx + Rp_{b'}(x) & x\in[0,b) \\
            -cx + R\frac{(1-\alpha)p_{b'}(x)+\alpha(x-b)}{1-\alpha + \alpha(x-b)} & x\in[b,b+1) \\
            -cx + (\alpha +(1-\alpha)p_{b'}(x))R & x\in[b+1,\infty)
        \end{cases}
    \end{equation*}
    It is easy to see that $U^{b',b}(x)$ is decreasing respectively in $[0,b)$ and $[b,\infty)$ regardless of the values of $b$ and $b'$. If $[b,b+1)\cap[b',b'+1)$ is empty, $p_{b'}(x)$ takes a constant value ($0$ or $1$) in the second line of the equation above, and it is easy to check that $U^{b',b}(x)$ decreases over $[b,b+1)$. If $[b,b+1)\cap[b',b'+1)$ is not empty, we have $U^{b',b}(x)=h(x)$ for $x\in[b,b+1)\cap[b',b'+1)$. If additionally, we assume $b'\leq b$, then $\overline{x}_h\leq b$ by the condition $\frac{R}{c}<1-\alpha$, implying that $U^{b',b}(x)$ decreases over $[b,b+1)\cap[b',b'+1)$. On the contrary, if we assume $b'>b$, then we observe
    \begin{equation*}
    \begin{aligned}
        b' - \overline{x}_h &= b'-b+\overline{\alpha} -\sqrt{\frac{\overline{\alpha}R}{\alpha c} + \frac{(1-\alpha)R}{\alpha c}(b'-b)} \\
        & > b'-b + \overline{\alpha} - \sqrt{\overline{\alpha}^2+\overline{\alpha}(1-\alpha)(b'-b)} \\
        &=\sqrt{\overline{\alpha}^2 + 2\overline{\alpha} (b'-b) + (b'-b)^2} \\ &\qquad - \sqrt{\overline{\alpha}^2+\overline{\alpha}(1-\alpha)(b'-b)} \\
        &> 0,
    \end{aligned}
    \end{equation*}
    where the first inequality follows from the condition $\frac{R}{c}<1-\alpha$ and the second condition follows from comparing the two terms in the subtraction. Thus, $\overline{x}_h<b'$ and $U^{b',b}(x)$ is still decreasing over $[b,b+1)\cap[b',b'+1)$. For any $x$ such that $x\in[b,b+1)$ but $x\not\in[b',b'+1)$, $p_{b'}(x)$ is a constant function and henceforth $U^{b',b}(x)$ is decreasing. We have shown that $U^{b',b}(x)$ is decreasing over $\mathbb{R}_+$, regardless of the values of $b$ and $b'$. Therefore, there are no audits that incentivize such vendors.

    When $\frac{R}{c}\geq \frac{1}{1-\alpha}$, the maximum utility of the vendor in the repeated audit starting from step 2 is:
\begin{equation*}
    \begin{aligned}   
    % &U^{p,*}(x)\;\Big|\;b\in\left[0,\frac{R}{c}-1\right) = \begin{cases}
    %     R - c(b+1) & x\in[0,b+1) \\
    %     R - cx & x\in[b+1,\infty)
    % \end{cases}. \\
    U^{p,*}(x) = \begin{cases}
        -cx & x\in\left(-\infty,b+1-\frac{R}{c}\right) \\
        R-c(b+1) & x\in\left[b+1-\frac{R}{c}, b+1\right) \\
        R - cx & x\in\left[b+1,\infty\right)
    \end{cases},
    \end{aligned}
\end{equation*}
where we have extended the domain to $(-\infty,\infty)$ for easier analysis. Then, we derive the utility of the vendor starting from the first stage:
\begin{align}\label{eq:impossibility-largeR}
    &U^{b',b}(x)=\\&\begin{cases}
        -cx + Rp_{b'}(x) & x\in\left(-\infty,\; b+1-\frac{R}{c}\right) \\
        \begin{aligned}
            -cx + (\alpha + (1-\alpha)p_{b'}(x))R& \\ + \alpha c(1-p_{b'}(x))(x-b-1)&
        \end{aligned}
         & x\in\big[b+1-\frac{R}{c},\;b+1\big) \\
        -cx + (\alpha + (1-\alpha)p_{b'}(x))R & x\in[b+1,\;\infty)
    \end{cases}\nonumber
\end{align}
The maximum vendor's utility can be calculated as $\max_{x\geq 0}\;U^{b',b}(x)$. We take the following result for granted as of now and defer the proof later to Lemma \ref{lemma:mono-ubb}: $U^{b',b}(x)$ is decreasing respectively over $(-\infty,b')$ and $[b'+1,\infty)$, and is increasing over $[b',b'+1)$.

So, the incentivizable investment (if it exists) must equal the local maximizer $b'+1$. We can rewrite the optimization problem \eqref{eq:dynamic-optimization} as
\begin{equation*}
    \begin{aligned}
        \max_{b',b\geq 0}\;&b'+1\\
        s.t.\;&U^{b',b}(b'+1)\geq 0 \\
    \end{aligned}
\end{equation*}
This is a simple linear program, and the solution is given by $b'=\frac{R}{c}-1$ and $b\in\mathbb{R}_+$. This implies that the vendors always invest enough to pass the audit in the first stage, making all subsequent stages useless.
\end{proof}

In the proof of Lemma \ref{lemma:dynamic-impossibility}, we utilized the following result.
\begin{lemma}\label{lemma:mono-ubb}
    Suppose $\frac{R}{c}\geq \frac{1}{1-\alpha}$. For any $b$ and $b'$, $U^{b',b}(x)$ is decreasing respectively over $(-\infty,b')$ and $[b'+1,\infty)$ and increasing over $[b',b'+1)$.
\end{lemma}
\begin{proof}:
    It is easy to verify that each piece of $U^{b',b}(x)$ (displayed in Eq. \eqref{eq:impossibility-largeR}) is decreasing for either $p_{b'}(x)\equiv 0$ (when $x\leq  b'$) or $p_{b'}(x)\equiv 1$ (when $x\geq b'+1$). (1) For any $x\in\left(-\infty,\; b+1-\frac{R}{c}\right)\cap[b',b'+1)$, we have $U^{b',b}(x)=-cx+R(x-b')=(R-c)x-Rb'$ increases in $x$. (2) For any $x\in\left[b+1-\frac{R}{c},\;b+1\right)\cap[b',b'+1)$, we plug in $p_{b'}(x)=x-b'$ and obtain
    \begin{equation*}
        U^{b',b}(x) = -\alpha cx^2 + \left[-c+(1-\alpha)R + \alpha c(b'+b+2)\right]x + K,
    \end{equation*}
    where $K$ is some irrelevant constant. The axis of symmetry of this parabola is
    \begin{equation*}
        \tilde{x} = 1+\frac{b'+b}{2} + \frac{1-\alpha}{2\alpha c}\left(R-\frac{c}{1-\alpha}\right).
    \end{equation*}
    Notice $\tilde{x} - \min\{b+1,b'+1\}\geq \frac{1}{2}(b'+b-2\min\{b',b\})\geq 0$, where the first inequality follows from $\frac{R}{c}\geq \frac{1}{1-\alpha}$. Thus, $U^{b',b}(x)$ is also increasing in $x$. (3) For any $x\in\left[b+1,\infty\right)\cap[b',b'+1)$, we have $U^{b',b}(x)=\left[R(1-\alpha)-c\right]x + (\alpha - (1-\alpha)b')R$ which is also increasing in $x$.
\end{proof}

\subsection{Complete Proofs of Proposition \ref{prop:dynamic-solution} and \ref{prop:dynamic-impossible-harder}}\label{app:dynamic-incentive}
As is shown by Lemma \ref{lemma:dynamic-impossibility}, we only examine the situation where $1-\alpha\leq \frac{R}{c} < \frac{1}{1-\alpha}$. 
In this case, the explicit formula for $U^{b,*}(x)$ can be obtained by solving $\max_{z\geq x}\;G^b(z)$:
\begin{equation*}
    U^{b,*}(x) = \begin{cases}
        -cx & x\in S_1 \\
        \left(\sqrt{\frac{R}{\alpha}}-\sqrt{\overline{\alpha}c}\right)^2-bc & x\in S_2 \\
         -cx + \frac{x-b}{1-\alpha+\alpha(x-b)}R& x\in S_3 \\
        -cx+R & x\in S_4
    \end{cases}
\end{equation*}
where 
\begin{equation*}
    \begin{aligned}
        &S_1:=\big(-\infty,\; b - \left(\sqrt{\frac{R}{\alpha c}}-\sqrt{\overline{\alpha}}\right)^2\big) \\
        &S_2:=\big[b - \left(\sqrt{\frac{R}{\alpha c}}-\sqrt{\overline{\alpha}}\right)^2,\;b-\overline{\alpha} + \sqrt{\frac{\overline{\alpha}R}{\alpha c}}\big) \\
        &S_3:=\big[b-\overline{\alpha} + \sqrt{\frac{\overline{\alpha}R}{\alpha c}},\; b+1\big) \\
        &S_4:=[b+1,\infty).
    \end{aligned}
\end{equation*}
It can be verified that the second and third regions are always non-empty. We place this expression into $U^{b',b}(x):=-(1-\alpha + \alpha p_{b'}(x))C_X(x) + p_{b'}(x)R + \alpha(1-p_{b'}(x))U^{b,*}(x)$ and obtain
\begin{equation}\label{eq:expr-ubb}
    U^{b',b}(x) = \begin{cases}
        -cx + Rp_{b'}(x) & x\in S_1 \\
        % \begin{aligned}
        %     -&(1-\alpha + \alpha p_{b'}(x))cx + Rp_{b'}(x) + \\ 
        %     &\; \alpha (1-p_{b'}(x))\big[\big(\sqrt{R/\alpha}-\sqrt{\overline{\alpha}c}\big)^2-bc\big]
        % \end{aligned}
        -cx + R + (1-p_{b'}(x))(\alpha cx + K)
        % \begin{aligned}
        % &-cx + R + (1-p_{b'}(x))
        % \big[\alpha cx + \\
        % &\;\;\;\;(1-\alpha) c - 2\sqrt{(1-\alpha)Rc} - \alpha bc\big]
        % \end{aligned}
        & x\in S_2 \\
        -cx + \frac{(1-\alpha)p_{b'}(x)+\alpha(x-b)}{1-\alpha + \alpha(x-b)}R & x\in S_3 \\
        -cx + (\alpha + (1-\alpha)p_{b'}(x))R & x\in S_4
    \end{cases},
\end{equation}
where $K:=(1-\alpha) c - 2\sqrt{(1-\alpha)Rc} - \alpha bc$. 

\begin{proposition}[Restatement of Proposition \ref{prop:dynamic-solution}]
    Suppose $1< \frac{R}{c}< \frac{1}{1-\alpha}$. The test functions characterized by the following entrance values constitute a solution to \eqref{eq:dynamic-optimization}:
    \begin{equation*}
        \overline{b} = \frac{R}{c} + \left(\sqrt{\frac{R}{\alpha c}}-\sqrt{\overline{\alpha}}\right)^2\quad\text{and}\quad \overline{b}' = \frac{R}{c} - 1
    \end{equation*}
    with the incentivizable investment $x(\overline{b},\overline{b}')=\frac{R}{c}$.
\end{proposition}
\begin{proof}
    Suppose $b'+1<b$. Then, $U^{b',b}(x) = (R-c)x - Rb'$ when $x\in[b',b'+1)$ and is increasing on $[b',b'+1)$. Also, since $p_{b'}(x)\equiv 1$ for $x\geq b'+1$, it is easy to verify that $U^{b',b}(x)$ is decreasing in each of $[b'+1,\infty)\cap S_1$, $S_2$, $S_3$, and $S_4$. For $x\in[0,b')$, $U^{b',b}(x)=-cx$ is also decreasing. The idea is to choose $b$ and $b'$ such that $b'+1<b$ and the vendor's maximum utility is attained at $b'+1$. Problem \eqref{eq:dynamic-optimization} restricted to this case is equivalent to
    \begin{equation*}
        \begin{aligned}
            \max_{b',\;b\geq 0}\;&\; b'+1 \\
            s.t.\;&\;U^{b',b}(b'+1) = R - c(1+b')\geq 0 \\
            &\;b'+1\leq L:=b-\Big(\sqrt{\frac{R}{\alpha c}}-\sqrt{\frac{1-\alpha}{\alpha}}\Big)^2.
        \end{aligned}
    \end{equation*}
    The first constraint satisfies both IC and VP because the other local maximum, in this case, is $x=0$ and $U^{b',b}(0)=0$. This is a linear program and can be easily solved. The proof is completed by observing that the obtained $x(\overline{b},\overline{b}')=\frac{R}{c}$ is indeed the global maximum as it equates to the maximum theoretical investment.
\end{proof}

\begin{proposition}[Restatement of Proposition \ref{prop:dynamic-impossible-harder}]
    Suppose we add a constraint $b+\varepsilon \leq b' \leq b+1$ to the optimization problem \eqref{eq:dynamic-optimization} with $\varepsilon>0$ emphasizing the minimum difference between the two tests. When $1-\alpha < \frac{R}{c} < \frac{1}{1-\alpha^2}$, the solution $(\overline{b},\;\overline{b}')$ to the modified optimization problem exists, 
    \begin{equation*}
        \overline{b} = \overline{\alpha}+\frac{R}{\alpha c} - 2\sqrt{\frac{\overline{\alpha}R}{\alpha c} + \frac{\overline{\alpha}R}{c}\varepsilon}\quad\text{and}\quad\overline{b}'=\overline{b} + \varepsilon
    \end{equation*}
    and the incentivizable investment is
    \begin{equation*}
        x(\overline{b},\overline{b}') = \overline{b}-\overline{\alpha} + \sqrt{\frac{\overline{\alpha}R}{\alpha c} + \frac{\overline{\alpha}R}{ c}\;\varepsilon},
    \end{equation*}
When $\frac{1}{1-\alpha^2}\leq R < \frac{1}{1-\alpha}$, the above audit is still optimal given that $\varepsilon$ is sufficiently small, i.e., $\varepsilon<\frac{c}{\alpha(1-\alpha)R}-\frac{1}{\alpha}$. Otherwise, the solution to the modified optimization problem, denoted alternatively as $(\Tilde{b},\Tilde{b}')$, is given by
\begin{equation*}
    \Tilde{b} = \frac{R - (1+\alpha)c}{\alpha c}\quad\text{and}\quad \Tilde{b}' = \frac{R}{\alpha c} + \frac{c}{\alpha (1-\alpha)R} - \frac{2+\alpha}{\alpha}
\end{equation*}
with the incentivizable investment $x(\Tilde{b},\Tilde{b}')=\frac{R-c}{\alpha c}$.
Furthermore, both $x(\overline{b},\overline{b}')$ and $x(\Tilde{b},\Tilde{b}')$ are less than $\frac{R-\sqrt{(1-\alpha)Rc}}{\alpha c}$, the incentivizable investment for static audits.
\end{proposition}
\begin{proof}
    We first show that when $b<b'\leq b+1$, the vendor's optimal investment can only take three values: $0$, $\overline{x}_h$, or $b+1$. Finding the incentivizable investment is sensible only when the vendor's investment takes the latter two values.
    % We show that, when $b<b'\leq b+1$, the nonzero local maximum of $U^{b',b}(\cdot)$ is $\overline{x}_h$ when $1-\alpha \leq \frac{R}{c} < \frac{1}{1-\alpha^2}$
    % and the corresponding local maximum value is $h(\overline{x}_h)$. 
    This follows readily from the following intermediate results and the monotonicity of $h(x)$.
    \begin{enumerate}[(a)]
        \item $U^{b',b}(x)$ decreases respectively over $S_1$ and $S_4$ regardless of the values of $b$ and $b'$.
        \item In $S_2$ and given $b'>b$, $U^{b',b}(x)$ increases over $\left[b',\;b - \overline{\alpha} + \sqrt{\frac{\overline{\alpha}R}{\alpha c}}\right)$ and decreases over $\left[b-\left(\sqrt{\frac{R}{\alpha c}}-\sqrt{\overline{\alpha}}\right)^2,\;\min\left\{b',\;b - \overline{\alpha} + \sqrt{\frac{\overline{\alpha}R}{\alpha c}}\right\}\right)$. (If the first interval is empty, then it implies $U^{b',b}(x)$ decreases monotonically over $S_2$.)
        \item Given $\frac{R}{c}\geq 1-\alpha$ and $b'>b$, we have $\overline{x}_h>b-\overline{\alpha}+\sqrt{\frac{\overline{\alpha}R}{\alpha c}+\frac{\overline{\alpha}R}{c}(b'-b)}$.
    \end{enumerate}
    For (a), $U^{b',b}(x)$ decreases over $S_1$ because $b'>b>b-\left(\sqrt{\frac{R}{\alpha c}}-\sqrt{\overline{\alpha}}\right)^2$ where the last term is the right boundary of $S_1$; and it decreases over $S_4$ because the coefficient of the linear term in $S_4$ is at most (not including) 0 due to $\frac{R}{c}<\frac{1}{1-\alpha}$. For (b), it is easy to verify that $U^{b',b}(x)$ is decreasing if $x\in S_2\cap [b',b'+1)^c$ and $U^{b',b}(x)=g(x)$ if $x\in S_2\cap [b',b'+1)$. In the latter case, since $b'>b$, we have $\overline{x}_g>b-\overline{\alpha}+\sqrt{\frac{\overline{\alpha}R}{\alpha c}}$, where the right hand side is the right boundary of $S_2$. Thus, $U^{b',b}$ is increasing over $S_2\cap[b',b'+1)$. The stated intervals are simplifications of $S_2\cap[b',b'+1)$ and $S_2\cap[b',b'+1)^c$ respectively by the observation $b'>b>b-\left(\sqrt{\frac{R}{\alpha c}}-\sqrt{\overline{\alpha}}\right)^2$.  For (c), $b'>b$ implies $\overline{x}_h>b-\overline{\alpha} + \sqrt{\frac{\overline{\alpha}R}{\alpha c}+ \frac{(1-\alpha)R}{\alpha c}(b'-b)}>b-\overline{\alpha}+\sqrt{\frac{\overline{\alpha}R}{\alpha c}}$.
    % and $\frac{R}{c}<\frac{1}{1-\alpha^2}$ with $b'\leq b+1$ implies $\overline{x}_h< b -\overline{\alpha} + \sqrt{\frac{1+\alpha(b'-b)}{\alpha^2(1+\alpha)}}\leq b+1$. Thus, we obtain $\overline{x}_h\in S_3$. 

    % Notice that (c) does not provide information on the relationship between $b'$ and $\overline{x}_h$. 
    When $\overline{x}_h<b'$, $U^{b',b}(x)$ is decreasing over $S_2$ and $S_3$. The reason is that for $S_2$, (c) implies $b'>\overline{x}_h\geq b-\overline{\alpha}+\sqrt{\frac{\overline{\alpha}R}{\alpha c}}$ and thus the first interval in (b) is empty; for $S_3$, $U^{b',b}(x)$ decreases in $S_3\cap[b',b'+1)^c$ because $p_{b'}(x)=0$ and also in $S_3\cap[b',b'+1)$ because the local maximizer is less than the left boundary of the interval, i.e., $\overline{x}_h<b'$. Therefore, the vendor cannot be further incentivized when $\overline{x}_h<b'$. 
    
    When $b'\leq \overline{x}_h \leq b+1$, (b) and (c) implies that $U^{b',b}(x)$ is increasing in $[b',\overline{x}_h)$ and decreasing outside this interval.
    Thus, the non-zero local maximum of $U^{b',b}(x)$ is attained at $\overline{x}_h$. The problem is equivalent to
    \begin{equation*}
        \begin{aligned}
            \max_{b,b'\geq 0}\;&\;\overline{x}_h \\
            s.t.\;&\; U^{b',b}(\overline{x}_h)=h(\overline{x}_h)\geq 0\\
            &\;b+\varepsilon\leq b'\leq b+1 \\
            &\;b'\leq \overline{x}_h \leq b+1
        \end{aligned}
    \end{equation*}
    where $\varepsilon> 0$ is a small constant used to exclude the solution $b=b'$. The first constraint is equivalent to
    \begin{equation*}
    \begin{aligned}
        h(\overline{x}_h)\geq 0 \iff 
        b-\overline{\alpha} - \frac{R}{\alpha c} + 2\sqrt{\frac{\overline{\alpha}R}{\alpha c} + \frac{\overline{\alpha}R}{ c}(b'-b)}\leq 0
        % \overline{x}_h \leq \frac{R}{\alpha c} - \sqrt{\frac{\overline{\alpha}R}{\alpha c} + \frac{\overline{\alpha}R}{ c}(b'-b)}
    \end{aligned}
    \end{equation*}
    and the third constraint is equivalent to
    \begin{equation*}
        \begin{aligned}
            b'\leq \overline{x}_h &\iff b'\leq b-\overline{\alpha} + \sqrt{\frac{\overline{\alpha}R}{\alpha c} + \frac{\overline{\alpha}R}{c}(b'-b)} \\
            \iff &(b'-b)^2 +\left(2\overline{\alpha} -\frac{\overline{\alpha}R}{ c}\right)(b'-b) + \overline{\alpha}^2-\frac{\overline{\alpha}R}{\alpha c}\leq 0 \\
            \overline{x}_h \leq b+ 1 &\iff b-\overline{\alpha} + \sqrt{\frac{\overline{\alpha}R}{\alpha c} + \frac{\overline{\alpha}R}{c}(b'-b)} \leq b+1 \\
            &\iff b'-b < \frac{c}{\alpha(1-\alpha)R} -\frac{1}{\alpha}
        \end{aligned}
    \end{equation*}
    Now, denote $\Delta:=b'-b$. The problem can be rewritten according to the above two observations as follows.
    \begin{equation}\label{eq:prob-1}
        \begin{aligned}
            \max_{b\geq 0, \Delta}\;&\;\overline{x}_h:=b-\overline{\alpha} + \sqrt{\frac{\overline{\alpha}R}{\alpha c}+ \frac{\overline{\alpha}R}{c}\Delta} \\
            s.t.\;&\; b-\overline{\alpha} - \frac{R}{\alpha c} + 2\sqrt{\frac{\overline{\alpha}R}{\alpha c} + \frac{\overline{\alpha}R}{ c}\Delta}\leq 0 \\
            &\;\varepsilon \leq \Delta \leq b+1 \\
            &\;\Delta^2+\left(2\overline{\alpha}-\frac{\overline{\alpha}R}{c}\right)\Delta + \overline{\alpha}^2 - \frac{\overline{\alpha}R}{\alpha c}\leq 0 \\
            &\;\Delta\leq \frac{c}{\alpha(1-\alpha)R} -\frac{1}{\alpha}
        \end{aligned}
    \end{equation}
    When $\frac{R}{c}=1-\alpha$, the first constraint implies $\Delta\leq \frac{c}{\overline{\alpha}R}\left(\overline{\alpha}-\frac{R}{\alpha c}\right)^2=0$, which contradicts with the second constraint. When $\frac{R}{c}>1-\alpha$, we can always choose $\varepsilon$ small enough s.t. the feasible set is non-empty, i.e., containing the feasible point $b=0$ and $\Delta=\varepsilon$. 
    Thus, this problem is feasible if and only if $\frac{R}{c}> 1-\alpha$ and $\varepsilon$ is sufficiently small. Observe that the first constraint is equivalent to 
    \begin{equation}\label{eq:xhbar-ubb}
        \overline{x}_h\leq \frac{R}{\alpha c} - \sqrt{\frac{\overline{\alpha}R}{\alpha c} + \frac{\overline{\alpha R}}{c}\Delta}.
    \end{equation}
    Observe that $\overline{x}_h$ is linear in $b$, and the second, third, and fourth constraints do not involve $b$. So, we can always choose $b$ for each $\Delta$ such that constraint \eqref{eq:xhbar-ubb} is active (attaining equality). Since the right-hand side of Eq. \eqref{eq:xhbar-ubb} is decreasing in $\Delta$, the solution is therefore $\overline{\Delta}=\varepsilon$ (that is $\overline{b}'=\overline{b}+\varepsilon$) and $\overline{b} = \overline{\alpha}+\frac{R}{\alpha c} - 2\sqrt{\frac{\overline{\alpha}R}{\alpha c} + \frac{\overline{\alpha}R}{c}\varepsilon}$, with optimal objective value $x(\overline{b},\overline{b}')=\frac{R}{\alpha c} - \sqrt{\frac{\overline{\alpha}R}{\alpha c} + \frac{\overline{\alpha R}}{c}\varepsilon}$. As $\varepsilon>0$, we obtain $x(\overline{b},\overline{b}')\leq \frac{R}{\alpha c}-\sqrt{\frac{\overline{\alpha}R}{\alpha c}}$ which is incentivizable investment of static audits.

    When $\overline{x}_h\geq b+1$, then the non-zero local maximum is attained at $b+1$. Thus, the optimization problem is equivalent to the following.
    \begin{equation}\label{eq:prob-2}
        \begin{aligned}
            \max_{b\geq 0,\Delta}\;&\;b+1 \\
            s.t.\;&\;U^{b',b}(x) = -c(b+1) + R\left(1-(1-\alpha)\Delta\right) \geq 0 \\
            \;&\overline{x}_h\geq b+ 1 \\
            \;&\varepsilon\leq \Delta\leq 1
        \end{aligned}
    \end{equation}
    where $\Delta:=b'-b$ and $\varepsilon>0$ is a helper variable to mark distinct tests as above. The second constraint can be simplified to $\Delta \geq \frac{c}{\alpha(1-\alpha)R}-\frac{1}{\alpha}$. This problem is feasible if and only if $\frac{R}{c}\geq \frac{1}{1-\alpha^2}$. Then, the above problem becomes a linear program, and the solution can be easily solved and given by $\overline{\Delta}=\frac{c}{\alpha(1-\alpha)R}-\frac{1}{\alpha},\;\varepsilon$ and $\overline{b}=\frac{R}{c}(1-(1-\alpha)\overline{\Delta})-1$. The optimal objective is therefore $x(\overline{b},\overline{b}')=\frac{R}{c}(1-(1-\alpha)\overline{\Delta})$. According to the expression of $\overline{\Delta}$, the objective is upper bounded by $x(\overline{b},\overline{b}')= \frac{R}{c}\left(1-\frac{c}{\alpha R}+\overline{\alpha}\right)= \frac{R- c}{\alpha c} \leq \frac{R}{\alpha c}-\sqrt{\frac{\overline{\alpha}R}{\alpha c}}$.

    Lastly, we have shown that problem \eqref{eq:prob-1} is feasible iff $\frac{R}{c}> 1-\alpha$, while problem \eqref{eq:prob-2} is feasible iff $\frac{R}{c}\geq \frac{1}{1-\alpha^2}$. Thus, for vendors whose ROI satisfies $1-\alpha<\frac{R}{c}< \frac{1}{1-\alpha^2}$, its incentivizable investment is obtained by \eqref{eq:prob-1}. For vendors whose ROI satisfies $\frac{1}{1-\alpha^2}\leq \frac{R}{c} < \frac{1}{1-\alpha}$, there are two potential incentivizable investments depending on the difference between $b$ and $b'$. It can be verified that when $\varepsilon$ is small enough ($\varepsilon<\frac{c}{\alpha(1-\alpha)R}-\frac{1}{\alpha}$ which is a natural assumption), the audit obtained from \eqref{eq:prob-1} induces higher incentivizable investment than \eqref{eq:prob-2}, meaning a less different two-step audit is preferred.
\end{proof}

\appendix[Intermediate Result for Lemma \ref{lemma:one-change}]

In the proof of Lemma \ref{lemma:one-change}, we construct the inequality by claiming $R-\alpha (C_X(x)+U_{m+1}^{A,*}(x))\geq 0$ for all $x$. We show a slightly generalized version of this inequality: $R-\alpha(C_X(x)+U_{0}^{A,*}(x))\geq 0$ for all $x$ and audit $A$.
To see the relationship between the latter result to the original statement, observe that $U_{m+1}^{A,*}=U_0^{Q,*}$ where the functions in $Q:=\{q_n\}_{n=0}^\infty$ are $(m+1)$-left-shifted from those in $A$: $q_n = p_{n+m+1}$.
The proof of this statement involves induction arguments and is presented as follows.
\begin{lemma}\label{lemma:bdd}
    Define $\mathcal{A}_k:=\{A=\{p_n\}_{n=0}^\infty\;|\;p_{k+i}=p_k,\;\forall i\geq 0\}$ as the set of audits whose test functions after step $k$ are fixed. 
    Then, $0\leq C_X(\cdot)+U_0^{A_k,*}(\cdot)\leq R$ for every $A_k\in\mathcal{A}_k$ and for every $k\geq 0$. As a consequence, we have $0\leq C_X(\cdot)+U_0^{A,*}(\cdot)\leq R$ for every arbitrary audit $A$.
\end{lemma}

\begin{proof}
    We prove this by induction over $k$. For the upper bound, observe that, as the base case, $\mathcal{A}_0$ denotes the set of static audits. Thus, for any $A_0\in\mathcal{A}_0$, $C_X(x)+U_0^{A_0,*}(x)=C_X(x)+W^{A_0}(x)=V^{A_0,*}(x)\leq R$ where $W^{A_0}(\cdot)$ and $V^{A_0,*}(\cdot)$ are the same functions as $W(\cdot)$ and $V^*(\cdot)$ defined in Section \ref{sec:continuation-investments}, with superscripts emphasizing their dependence on the audit $A_0$. The last inequality follows from the definition of the value function. 
    % Notice that $p_0$ can be arbitrary.
    
    For the hypothesis, assume $C_X(\cdot)+U_0^{A_k,*}(\cdot)\leq R$ holds for all $A_k\in\mathcal{A}_k$ and all $k = 0,1,\dots, m-1$. For any $A_m:=\{p_{m,n}\}_{n=1}^\infty\in\mathcal{A}_m$, construct another audit $A_{m-1}:=\{p_{m-1,n}\}_{n=0}^\infty$ by shifting $A_m$ to the left by one, i.e., $p_{m-1,n}=p_{m,n+1}$ for all $n\geq0$. Then, 
    \begin{equation}\label{eq:u0am}
        \begin{aligned}
            U_0^{A_{m},*}(x) &= \max_{z\geq x}\;-C_X(z) + R\cdot p_0(z) \\
            &\qquad + \alpha (1-p_0(z))\big[U^{A_m,*}_1(z)+C_X(z)\big] \\
            &= \max_{z\geq x}\;-C_X(z) + R\cdot p_0(z) \\
            &\qquad + \alpha (1-p_0(z))\big[U^{A_{m-1},*}_0(z)+C_X(z)\big] \\
            &\leq \max_{z\geq x}\;-C_X(z) + R\cdot \big[ \alpha + (1-\alpha)p_{m-1,0}(z)\big],
        \end{aligned}
    \end{equation}
    where the last inequality is justified by noticing that $A_{m-1}\in\mathcal{A}_{m-1}$. Thus, $C_X(x)+U_0^{A_m,*}(x)\leq \max_{z\geq x}\;\big[C_X(x)-C_X(z)\big] + R\cdot \big[ \alpha + (1-\alpha)p_{m-1,0}\big]\leq \max_{z\geq x}\; R\cdot \big[ \alpha + (1-\alpha)p_{m-1,0}\big]\leq R$ since $C_X(\cdot)$ is increasing and $p_{m-1,0}\in[0,1]$. 

    For the lower bound, let $A_0\in\mathcal{A}_0$ and the static audit function be $p_0$. Observe the base case $C_X(x)+U_0^{A_0,*}(x)\geq \frac{R\cdot p_0(x)}{1-\alpha + \alpha p_0(x)}\geq 0$ by Lemma \ref{lemma:explicit-w}. For the hypothesis, assume $C_X(\cdot)+U_0^{A_k,*}(\cdot)\geq 0$ for all $A_k\in\mathcal{A}_k$ and for all $k=0,1,\dots, m-1$. Let $A_{m-1}$ be constructed in the same way as above. Thus, 
    \begin{equation*}
        \begin{aligned}
            &C_X(x) + U_0^{A_m,*}(x) \\
            \geq &R\cdot p_0(x) + \alpha (1-p_0(x))\big[U_0^{A_{m-1},*}(x)+C_X(x)\big]\\
            \geq &R\cdot p_0(x)\geq 0,
        \end{aligned}
    \end{equation*}
    where the first inequality is obtained by plugging $z=x$ into the expression of $U_0^{A_m,*}$ in Eq. \eqref{eq:u0am} and the second inequality comes from the hypothesis.
    
    The generalization to arbitrary audit $A$ can be shown by constructing $A_k$'s s.t. the first $k$ test functions are the same as $A$, and then take $k$ to $\infty$.
\end{proof}

\end{document}